\documentclass[twocolumn]{aastex63}

\usepackage[T1]{fontenc}
\usepackage{longtable}
\usepackage{threeparttable}
\usepackage{color,epsfig,graphics,graphicx,rotating,comment,amssymb,stackengine}
\usepackage{natbib,lipsum,atbegshi}
\usepackage{sidecap}
\usepackage{ulem}
\usepackage{enumitem}
\definecolor{brown}{rgb}{0.6,0.4,0.2}
\definecolor{purple}{rgb}{0.5,0,0.5}
\definecolor{pinegreen}{HTML}{008B72}
\shortauthors{Ravi et al.}

\def\one{{\,\sc i}}
\def\two{{\,\sc ii}}

\newcommand{\kms}{km\,s$^{-1}$}

\newcommand{\mic}{$\mu$m}

\shorttitle{Type Ic SN 2021krf}

\begin{document}

\title{Near-Infrared and Optical Observations of Type Ic SN 2021krf: Luminous Late-time Emission and Dust Formation} 

\author{Aravind P. Ravi}\affil{Department of Physics, University of Texas at Arlington, Box 19059,
Arlington, TX 76019, USA}
\author{Jeonghee Rho}\affil{SETI Institute, 339 Bernardo Ave., Ste. 200, Mountain View, CA 94043, USA}
\author{Sangwook Park}\affil{Department of Physics, University of Texas at Arlington, Box 19059,
Arlington, TX 76019, USA}
\author{Seong Hyun Park}\affil{Department of Physics and Astronomy, Seoul National University, Gwanak-ro 1, Gwanak-gu, Seoul, 08826, South Korea}
\author{Sung-Chul Yoon}\affil{Department of Physics and Astronomy, Seoul National University, Gwanak-ro 1, Gwanak-gu, Seoul, 08826, South Korea}
\author{T. R. Geballe}\affil{Gemini Observatory/NSF's National Optical-Infrared Astronomy Research Laboratory, 670 N. Aohoku Place, Hilo, HI, 96720, USA}
\author{Jozsef Vink\'o}\affil{CSFK Konkoly Observatory, Konkoly-Thege M. ut 15-17, Budapest, 1121, Hungary}\affil{Department of Optics and Quantum Electronics, University of Szeged, D\'om t\'er 9, Szeged, 6720 Hungary}\affil{ELTE E\"otv\"os Lor\'and University, Institute of Physics, P\'azmany P\'eter s\'et\'any 1/A, Budapest, 1117, Hungary}
\author{Samaporn Tinyanont}\affil{Department of Astronomy and Astrophysics, University of California, Santa Cruz, CA 95064, USA}
\author{K. Azalee Bostroem}\thanks{LSSTC Catalyst Fellow}\affil{Steward Observatory, University of Arizona, 933 North Cherry Avenue, Tucson, AZ 85721-0065, USA}
\author{Jamison Burke}\affil{Las Cumbres Observatory, 6740 Cortona Drive, Suite 102, Goleta, CA 93117-5575, USA}\affil{Department of Physics, University of California, Santa Barbara, Broida Hall, Mail Code 9520, Santa Barbara, CA 93106-9530, USA}
\author{Daichi Hiramatsu}\affil{Center for Astrophysics | Harvard \& Smithsonian, 60 Garden Street, Cambridge, MA 02138-1516, USA}
\affil{The NSF AI Institute for Artificial Intelligence and Fundamental Interactions, MIT, Cambridge, MA 02139, USA}
\author{D. Andrew Howell}\affil{Las Cumbres Observatory, 6740 Cortona Drive, Suite 102, Goleta, CA 93117-5575, USA}\affil{Department of Physics, University of California, Santa Barbara, Broida Hall, Mail Code 9520, Santa Barbara, CA 93106-9530, USA}
\author{Curtis McCully}\affil{Las Cumbres Observatory, 6740 Cortona Drive, Suite 102, Goleta, CA 93117-5575, USA}
\author{Megan Newsome}\affil{Las Cumbres Observatory, 6740 Cortona Drive, Suite 102, Goleta, CA 93117-5575, USA}\affil{Department of Physics, University of California, Santa Barbara, Broida Hall, Mail Code 9520, Santa Barbara, CA 93106-9530, USA}
\author{Estefania Padilla Gonzalez}\affil{Las Cumbres Observatory, 6740 Cortona Drive, Suite 102, Goleta, CA 93117-5575, USA}\affil{Department of Physics, University of California, Santa Barbara, Broida Hall, Mail Code 9520, Santa Barbara, CA 93106-9530, USA}
\author{Craig Pellegrino}\affil{Las Cumbres Observatory, 6740 Cortona Drive, Suite 102, Goleta, CA 93117-5575, USA}\affil{Department of Physics, University of California, Santa Barbara, Broida Hall, Mail Code 9520, Santa Barbara, CA 93106-9530, USA}
\author{Regis Cartier}\affil{Gemini Observatory, NSF’s National Optical-Infrared Astronomy Research Laboratory, Casilla 603, La Serena, Chile }
\author{Tyler Pritchard}\affil{Center for Cosmology and Particle Physics, New York University, 726 Broadway, NY, NY 11201, USA}
\author{Morten Andersen}\affil{Gemini Observatory/NSF's National Optical-Infrared Astronomy Research Laboratory, Casilla 603, La Serena, Chile}\affil{European Southern Observatory, Karl-Schwarzschild-Strasse 2, D-85748 Garching bei M{\"u}nchen, Germany}
\author{Sergey Blinnikov}\affil{17 NRC Kurchatov Institute, Moscow, 123182; Sternberg Astronomical Institute (SAI) of Lomonosov MSU, Moscow, 119234; 
Dukhov Research Institute of Automatics (VNIIA), Moscow 127055}
\author{Yize Dong}\affil{Department of Physics  and Astronomy, University of California, 1 Shields Avenue, Davis, CA 95616-5270, USA}
\author{Peter Blanchard}\affil{Center for Interdisciplinary Exploration and Research in Astrophysics and Department of Physics and Astronomy, Northwestern University, 1800 Sherman Avenue, 8th Floor, Evanston, IL 60201, USA}
\author{Charles D. Kilpatrick}\affil{Center for Interdisciplinary Exploration and Research in Astrophysics and Department of Physics and Astronomy, Northwestern University, 1800 Sherman Avenue, 8th Floor, Evanston, IL 60201, USA}
\author{Peter Hoeflich}\affil{Department of Physics, Florida State University, 77 Chieftan Way, Tallahassee, FL 32306, USA}
\author{Stefano Valenti}\affil{Department of Physics  and Astronomy, University of California, 1 Shields Avenue, Davis, CA 95616-5270, USA}
\author{Alexei V. Filippenko}\affil{Department of Astronomy, University of California, Berkeley, CA 94720-3411, USA}
\author{Nicholas B. Suntzeff}\affil{George P. and Cynthia Woods Mitchell Institute for Fundamental Physics and Astronomy, Department of Physics and Astronomy, Texas A$\&$M University, College
Station, TX 77843, USA} 
\author{Ji Yeon Seok}\affil{Korea Astronomy and Space Science Institute, Daejeon, 305–348, Republic of Korea} 
\author{R. K\"onyves-T\'oth}\affil{CSFK Konkoly Observatory, Konkoly-Thege M. ut 15-17, Budapest, 1121, Hungary}
\affiliation{Department of Experimental Physics, Institute of Physics, University of Szeged, D\'om t\'er 9, Szeged, 6720 Hungary}
\affiliation{ELTE Eötvös Loránd University, Gothard Astrophysical Observatory, Szombathely, Hungary}
\author{Ryan J. Foley}\affil{Department of Astronomy and Astrophysics, University of California, Santa Cruz, CA 95064, USA}
\author{Matthew R. Siebert}\affil{Space Telescope Science Institute, 3700 San Martin Drive, Baltimore, MD 21218, USA}
\author{David O. Jones}\affil{Gemini Observatory/NSF's National Optical-Infrared Astronomy Research Laboratory, 670 N. Aohoku Place, Hilo, HI, 96720, USA}

\begin{abstract}
We present near-infrared (NIR) and optical observations of the Type Ic supernova (SN~Ic) SN 2021krf obtained between days 13 and 259 at several ground-based telescopes. The NIR spectrum at day 68 exhibits a rising $K$-band continuum flux density longward of $\sim 2.0$\,$\mu$m, and a late-time optical spectrum at day 259 shows strong [O~I] 6300 and 6364\,\AA\ emission-line asymmetry, both indicating the presence of dust, likely formed in the SN ejecta. We estimate a carbon-grain dust mass of $\sim 2 \times 10^{-5}$\,M$_{\odot}$ and a dust temperature of $\sim 900$--1200\,K associated with this rising continuum and suggest the dust has formed in SN ejecta.  Utilizing the one-dimensional multigroup radiation hydrodynamics code STELLA, we present two degenerate progenitor solutions for SN 2021krf, characterized by C-O star masses of 3.93 and 5.74\,M$_\odot$, but with the same best-fit $^{56}$Ni mass of 0.11\,M$_\odot$ for early times (0--70 days). At late times (70--300 days), optical light curves of SN 2021krf decline substantially more slowly than that expected from $^{56}$Co radioactive decay. Lack of H and He lines in the late-time SN spectrum suggests the absence of significant interaction of the ejecta with the circumstellar medium. We reproduce the entire bolometric light curve with a combination of radioactive decay and an additional powering source in the form of a central engine of a millisecond pulsar with a magnetic field smaller than that of a typical magnetar.

\end{abstract}


\section{Introduction}

The significance of core-collapse supernovae (CCSNe) as major dust factories in the early universe has been the subject of a long-standing debate. Dust formation in the early universe is implied by the large amount of dust observed in high-redshift ($z$) galaxies \citep{isaak02, bertoldi03, laporte17, fudamoto21}. While most of the dust observed in present-day galaxies is considered to originate from stellar winds of asymptotic giant branch (AGB) stars \citep[e.g.,][]{gehrz89, draine09p, matsuura09}, such stars would not be significant contributors to the high-$z$ dust, as the universe was too young for AGB stars to have formed \citep{morgan03, dwek08}. On the other hand, CCSNe can occur several million years after their massive progenitor stars form. Young remnants of these types of supernova (SN) explosions have been confirmed to have formed dust in their ejecta, such as Cas~A \citep{rho08, rho12, delooze17, niculescu-duvaz21}, SN~1987A \citep[e.g.,][]{suntzeff90, matsuura15, wesson15, bevan16}, G54.1+0.3 \citep{rho18}, and the Crab Nebula \citep{gomez13}. In addition, based on their models of observed late-time asymmetries in optical line profiles of a large set of CCSNe, \cite{niculescu-duvaz22} concluded that dust formed in their ejecta. This suggests that CCSNe can be a viable source of significant dust formation in the early Universe.

CCSNe are classified as Type Ic when their spectra do not obviously exhibit either H and He spectral lines \citep[e.g.,][]{filippenko1997optical, galYam17, williamson19}. These SNe are thought to be explosions of massive stars that have lost their H envelope and most, if not all, of their He envelope. Considerable controversy exists over the interpretation of the absence in some CCSNe of optical He lines, especially regarding whether that is evidence for He deficiency in the ejecta \citep{dessart11, hachinger12}. There are potentially promising He lines in the near-infrared (NIR), and thus observing both the optical and NIR spectra of SNe~Ic for He signatures is crucial for understanding the properties of stripped-envelope CCSN progenitors.

As stripped-envelope SNe (Type Ib/c) have lost most of their H/He envelopes, their expanding ejecta are of lesser density than those of typical Type II SNe that have well-known dust signatures \citep[see][for a review]{sarangi18}. While having lower density ejecta could facilitate faster cooling and earlier dust formation, it might also reduce dust formation. Among all types of dust signatures observed among CCSNe, only a few are for Type Ib/c. The first report of dust formation in an SN~Ib was SN 1990I \citep{elmhamdi04}. Early dust-formation signatures were observed $\sim 50$ days after explosion in the Type Ibn SN 2006jc \citep{nozawa08, smith08, diCarlo08}. In SN 1990I, dust was suggested to originate in the SN ejecta while the dust in SN 2006jc was suggested to originate from the dense shell formed in the post-shock circumstellar material (CSM). {\it Spitzer} mid-infrared (MIR) continuum observations of Type Ic SN 2005at and SN 2007gr showed the presence of dust at significantly later epochs \citep{kankare14}. The sources of dust emission in both these SNe were suggested to be from IR echoes the in pre-existing dust and formation of dust in the expanding ejecta. NIR observations of the Type Ic SN 2020oi showed a rising continuum at day 63 which was suggested to originate from the newly formed dust in its ejecta \citep{rho21}.

SNe~Ic are generally believed to be powered by radioactive decay of $^{56}$Ni and its decay product, $^{56}$Co \citep{colgate69}. For a typical SN~Ic, the decay of luminosity at late times ($t > 100$\,days) is expected to be at least as rapid as the characteristic decay of $^{56}$Co (see \citealt{anderson19} for a discussion of stripped-envelope SN nickel masses). The late-time luminosity of SNe~Ic could be greater than the radioactive decay rate, if the SN had additional power sources such as late-time interactions with CSM or interstellar material (ISM), and/or energy input from a central engine \citep[e.g.,][who noted late-time CSM interactions and energy from a central engine in SN 2010mb and iPTF15dtg, respectively]{benami14, taddia19}.

Rapidly rotating neutron stars are believed to be remnants of CCSNe. The observed light curves of some of these SNe could be affected by the spindown of remnant neutron stars as their rotational energy is released in the form of relativistic magnetized winds \citep[e.g.,][]{kasen10magnetar, dessart12b}. Such a rapidly spinning newborn magnetar was first invoked to explain the peculiar evolution of the Type Ib SN 2005bf \citep{maeda07}. Central engine powering models have now been fit to large samples of Type I superluminous supernovae (SLSNe-I), which are H-poor \citep[see][for a review]{gal-yam19}. As SLSNe-I and SNe~Ic have similarities in spectroscopic evolution, especially at late times, the central engine might also be an important contributor in the light curves of some SNe~Ic. The effects of such a central engine may be negligible in the early-epoch SN light curves when the energy from radioactive decay of $^{56}$Ni would dominate the SN luminosity. However, over longer times as the radioactive decay powers down, their contributions may become significant, affecting the evolution of their late-time light curves \citep[e.g.,][]{kotera13}.

In this work, we report the results of our observations of the recently discovered Type Ic SN 2021krf based on our multi-epoch NIR spectroscopy using the Gemini North Telescope and the NASA Infrared Telescope Facility (IRTF), and optical photometry and spectroscopy from the Las Cumbres Observatory (LCO), the Keck-II 10\,m telescope, the Southern Astrophysical Research Telescope (SOAR), and the 3\,m Shane telescope at Lick Observatory. We find a rising NIR continuum due to emission from warm dust and a probable detection at day 68 of CO overtone-band emission, and discuss the implications of these for dust formation in SNe. We also present and discuss our spectrophotometric data spanning the first $\sim 300$\,days after the SN explosion. We find late-time ($t > 200$\,days) flux excesses above those expected from radioactive decay in the light curves of SN 2021krf. We consider several scenarios for additional power sources.

In Section \ref{sec:2}, we describe our observations, and in Section \ref{sec:3}, we present our results and analysis of the optical photometry and (11 sets of) spectroscopy spanning from the explosion to 350 days, and NIR spectroscopy obtained at 13, 43, and 68 days. Section \ref{sec:4} discusses the origin of the dust emission, optical spectral modeling, dynamic motion of SN ejecta, and hydrodynamic modeling of radioactive decay, and whether additional input energy from a magnetar is required to fit the bolometric light curve. Our conclusions are presented in Section \ref{sec:5}.

\section{Observations} \label{sec:2}


SN 2021krf (ZTF21aaxtctv) was first detected by the Zwicky Transient Facility \citep[ZTF;][]{bellm18, masci18, graham19} on 2021 April 30 (MJD 59334) at $m_{g} = 18.0035$\,mag \citep{munoz-arancibia2021}. Here we assume that the explosion date of SN 2021krf ($t_{0}$) was 2021 April 26 (MJD $59330 \pm 4.5$), which is the midpoint between the ZTF first discovery date (MJD 59334) and its last date of nondetection, 2021 April 21 (MJD 59325). SN 2021krf is located in the nearby galaxy 2MASX J12511712+0031138 at a distance of $\sim 65$\,Mpc based on NED \citep[][assuming H$_{0}$ = 67.8 km s$^{-1}$ Mpc$^{-1}$, ${\Omega}_{m} = 0.308$, ${\Omega}_{\rm vac} =0.692$]{NED_IPAC}\footnote{http://ned.ipac.caltech.edu/} with redshift $z = 0.01355$. Based on spectral data from the New Technology Telescope (NTT), it was classified as a Type Ic SN \citep{paraskeva}. We observed SN 2021krf at optical and NIR wavelengths as summarized in Table~\ref{SN21krfSpec}.

\subsection{Spectroscopy} \label{sec:2.1}

We obtained NIR (0.8--2.5\,${\mu}$m) spectra of SN 2021krf with the Gemini Near-Infrared Spectrograph (GNIRS) on the 8.1\,m Frederic C. Gilett Gemini North telescope, on 2021 June 8 and 2021 July 3 (UTC dates are used throughout this paper), as part of observing program GN-2021A-Q-126. Exposure times associated with these epochs were $6 \times 300$\,s and $4 \times 300$\,s, respectively. 
Unfortunately, the total integration time for the observation on 2021 July 3 was only 20\,min because that was the time left in our program. The short exposure produced a relatively low signal-to-noise ratio (SNR) spectrum above 2.3\,\mic. We configured GNIRS in the cross-dispersed mode, utilizing a 32 line mm$^{-1}$ grating and a 0\farcs45-wide slit to achieve a spectral resolution of $R \equiv \lambda/\Delta \lambda \approx 1200$ (250\,km\,s$^{-1}$). Standard stare/nod-along-slit mode (ABBA) with a nod angle of $3^{\prime\prime}$ was used for the observations. To minimize differences in the airmass, we used nearby early type-A dwarf stars, observed either just before or after SN 2021krf, as telluric and flux standards.

We performed data reduction utilizing both of the GNIRS cross-dispersed reduction pipeline \citep{cooke2005} and a manual, order-by-order reduction for the shortest-wavelength orders. We used manual reduction with standard IRAF \citep{tody86} and Figaro \citep{shortridge92} tools for flatfielding, spike removal, rectification of spectral images, extraction, wavelength calibration, and removal of hydrogen absorption lines in the spectra of the standard stars. For order-by-order reduction, spectral segments covering different orders were stitched together after small scaling factors were applied, to produce final continuous spectra between 0.81 and 2.52\,${\mu}$m. As an additional check, similar spectra were also obtained with {\tt XDGNIRS}, a PyRAF-based data-reduction pipeline \citep{mason15}. The standard ABBA method was used to perform sky subtraction and combining with the two-dimensional (2D) data, before final 1D spectral extraction. We note that the observed spectra in the 1.35--1.45\,${\mu}$m and 1.80--1.95\,${\mu}$m bands are affected by the relatively low atmospheric transmission, warranting caution for the reliability at these wavelengths.

We observed SN 2021krf with the short cross-dispersion (SXD) mode of the SpeX spectrograph \citep{rayner03} on the NASA InfraRed Facility Telescope (IRTF) on 2021 May 9. In this mode with the 0\farcs8 slit, the spectral resolution is $R \approx 1000$. Similar to the GNIRS observations, the SN was observed in an ABBA dithering pattern with an A0V star observed immediately before and the associated flatfield- and comparison-lamp observations observed after. We reduced the data using \texttt{spextool} \citep{cushing04}, which performed flatfielding, wavelength calibration, background subtraction, and spectral extraction. We then performed telluric correction using \texttt{xtellcor} \citep{vacca03}.

\begin{table*} 
\caption{NIR and Optical Spectroscopy of SN 2021krf}
\vspace{0.1cm}
\hskip-2.0cm
\begin{tabular}{lcccccc} \toprule
{Date} & {MJD} &  {Day} & {Telescope--Instrument} & {Total Exposure Time} & {Slit Width} & {Resolution $R$} \\
       &       &        &                         & {(s)}                 & {(arcsec)} & {($\lambda$/$\Delta \lambda$)}\\
\hline \hline
2021 April 26 & $59330 \pm 4.5$ & 0 (= $t_{0}$)$^{a}$ & ... & ... & ... &  ...\\
2021 May 8 & 59342 & 12 & 2\,m FTN--FLOYDS & 2700 & 2.0 & $\sim$500\\
\textbf{2021 May 9}$^{b}$ & 59343 & 13 & \textbf{3.2\,m IRTF-SpeX} & 2400 &  0.8 & \textbf{$\sim$1000} \\
2021 May 10 & 59344 & 14 & 3\,m Shane--Kast & 1200 & 2.0 & $\sim$700\\
2021 May 11 & 59345 & 15 & 2\,m FTN--FLOYDS & 2700 & 2.0 & $\sim$500\\
2021 May 17 & 59351 & 21 & 2\,m FTS--FLOYDS & 2700 & 2.0 & $\sim$400\\
2021 May 26 & 59360 & 30 & 2\,m FTS--FLOYDS & 2700 & 2.0 & $\sim$400\\
2021 June 4 & 59369 & 39 & 3\,m Shane--Kast & 1800 & 2.0 & $\sim$700\\
2021 June 5 & 59370 & 40 & 2\,m FTS--FLOYDS & 2700 & 2.0 & $\sim$400\\
\textbf{2021 June 8} & 59373 & 43 & \textbf{8.1\,m Gemini--GNIRS}$^{c}$ & 1800 & 0.45 & \textbf{$\sim$1200} \\
2021 June 13 & 59378 & 48 & 2\,m FTS --FLOYDS & 2700 & 2.0 & $\sim$400\\
2021 June 21 & 59386 & 56 & 2\,m FTS--FLOYDS & 2700 & 2.0 & $\sim$400\\
\textbf{2021 July 3} & 59398 & 68 & \textbf{8.1\,m Gemini--GNIRS} & 1200 & 0.45 & \textbf{$\sim$1200} \\
2021 July 6 & 59401 & 71 & 2\,m FTS--FLOYDS & 3600 & 2.0 & $\sim$400\\
2021 July 12 & 59407 & 77 & 4.1\,m SOAR--Goodman & 3600 & 1.0 & $\sim$1200 \\
2022 January 10 & 59589 & 259 & 10\,m Keck--DEIMOS & 3000 & 1.0 & $\sim$1600\\
\hline
\end{tabular}
\vspace{0.2cm}
{\renewcommand\labelitemi{}
\begin{itemize}[leftmargin=*, noitemsep]
    \item $^{a}$The estimated explosion date, 2021 April 26, is taken to be the middle point between the last nondetection reported by ZTF (2021 April 21) and the first detection (2021 April 30).
    \item $^{b}$NIR observations are marked in bold.
    \item $^{c}$Spectrophotometry of GNIRS spectra on 2020608 and 20210703 yield the following approximate magnitudes: $J$, $H$, $K = 16.3$, 16.0, 16.1, and 16.8, 16.4, 16.5\,mag, respectively.
\end{itemize}
}
\label{SN21krfSpec}%
\end{table*}%

We obtained 8 sets of optical spectra at the LCO with the FLOYDS spectrographs mounted on the 2\,m Faulkes Telescope North (FTN) at Haleakala (USA) and the identical 2\,m Faulkes Telescope South (FTS) at Siding Spring (Australia), through the Global Supernova Project \citep{howell19}, between 2021 May 8 and 2021 July 6. A 2$^{\prime \prime}$-wide slit was placed on the target at the parallactic angle \citep{filippenko82}. We extracted, reduced, and calibrated 1D spectra following standard procedures using the
FLOYDS pipeline \citep{valenti14}.

We obtained optical spectra with the Kast spectrograph \citep{miller1993lick} on the 3\,m Shane telescope at Lick Observatory on 2021 May 10 and 2021 June 4. The spectra were reduced using a custom data-reduction pipeline based on the Image Reduction and Analysis Facility (IRAF) \citep{tody86}.\footnote{The pipeline is publicly accessible at \url{https://github.com/msiebert1/UCSC_spectral_pipeline}.}
The pipeline performed flatfield correction using observations of a flatfield lamp.
The instrument response function was derived using observations of spectroscopic standard stars observed on the same night. The 2D spectra were extracted using the optimal extraction algorithm \citep{horne1986}.

We obtained an optical spectrum with the Goodman spectrograph \citep{clemens04} mounted on SOAR telescope on 2021 July 7. We reduced the Goodman spectrum following the usual steps making use of custom IRAF reduction scripts. These steps included bias subtraction, flatfielding, wavelength calibrations using Hg-Ar-Ne lamps, optimal extraction, and flux calibration using a sensitivity curve from a flux standard star observed on the same night as the SN spectrum.

We observed SN 2021krf with the Deep Imaging Multi-Object Spectrograph \citep[DEIMOS;][]{faber03} on the Keck-II 10\,m telescope on 2022 January 9.  We used the 600ZD grating, GG455 order-blocking filter, and 1\arcsec\ slit, integrating for $2 \times 1500$\,s. The approximate airmass was 1.27 during the observation, and we aligned the instrument to the parallactic angle.  We also observed the spectrophotometric standard Feige\,110 on the same night and in the same instrumental setup, which was used to derive the sensitivity function and flux calibration described below.

All DEIMOS data reductions were done with {\tt pypeit} \citep{prochaska2020}, which performs image-level calibration using the DEIMOS overscan region for bias correction and flatfielding using dome-flat frames, sky-line subtraction, and trace fitting and extraction.  We then derived a sensitivity function using the extracted standard-star spectrum and applied this to the SN 2021krf spectra.  Finally, we coadded the calibrated 1D spectra.

\subsection{Photometry} \label{sec:2.2}


We performed optical photometry ($U$, $B$, $g$, $V$, $r$, and $i$ filters) of SN 2021krf with follow-up LCO observations utilizing a world-wide network of telescopes under the Global Supernova Project \citep{howell19}. Point-spread-function (PSF) fitting for the images was performed utilizing a PyRAF-based photometric reduction pipeline, lcogtsnpipe\footnote{https://github.com/LCOGT/lcogtsnpipe} \citep{valenti16}. When the pre-SN images are not available from the LCO network in $g$, $r$, and $i$, we used the $gri$-band Sloan Digital Sky Survey (SDSS; SDSS Collaboration 2017) templates for image subtraction using PyZOGY \citep{guevel17}, an implementation in Python of the
subtraction algorithm described by \cite{zackay16}. The subtracted $gri$ images utilizing SDSS templates were calibrated to AB magnitudes \citep{oke1983}. For $U$, $B$, and $V$, LCO images taken respectively on 2022 January 14 (MJD = 59593), 2022 March 10 (MJD = 59648), and 2022 February 17 (MJD = 59627) were chosen as templates for difference imaging. The subtracted $UBV$ data were calibrated to Vega magnitudes. We obtained additional photometric data using the $gr$-band public ZTF data. We used the same pre-SN $gr$ images from SDSS as templates for ZTF difference imaging.

\section{Results} \label{sec:3}

\begin{figure}

\includegraphics[width=0.5\textwidth]{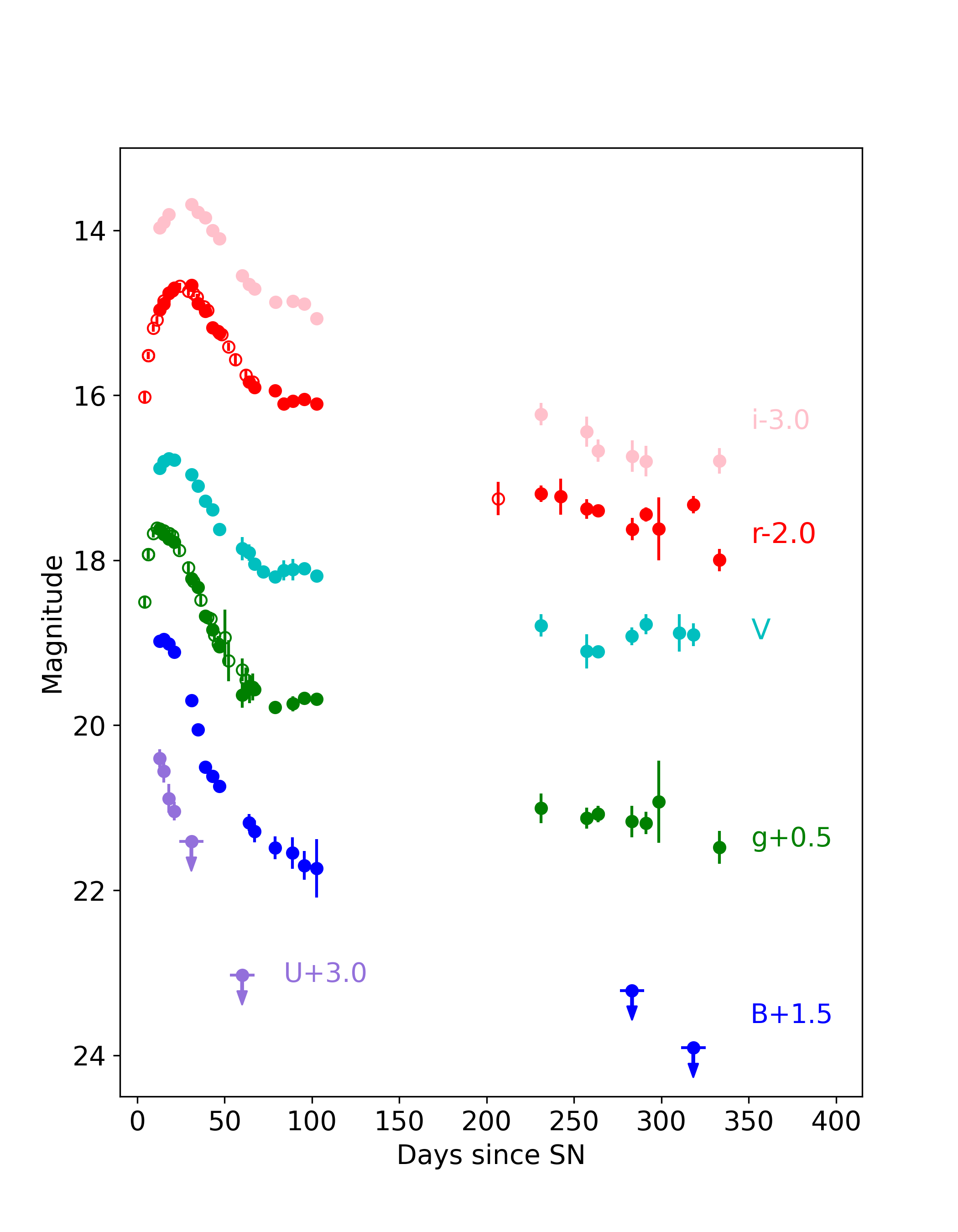}
\caption{Multicolor light curves of SN 2021krf combining LCO (filled circles) and ZTF (open circles). The explosion date of 2021 April 26 (MJD = 59330) is used as day 0 ($t_0$). The magnitudes are scaled (as labeled in the right-hand side of the plot) for the purposes of display. The final two $U$ and  $B$ measurements are upper limits.}

\label{Flightcurves21krf}
\end{figure}

\subsection{Optical Light Curves and Explosion Properties} \label{sec:3.1}

The optical light curves of SN 2021krf are shown in Figure \ref{Flightcurves21krf}. Based on ZTF follow-up observations of SN 2021krf, the $g$ and $r$ light curves gradually rise over the first $\sim 10$--20 days \citep{prentice16}. The $r$ light curve peaks at 24.21 days (MJD = 59354.21) after $t_{0}$, with an observed magnitude of 16.67 mag. 
Because of a gap between ZTF and LCO start dates, the $UBVi$ light curves that start $\sim$12 days after the SN rise only marginally before peaking. For example, the $V$ light curve peaks at 20.17 days (MJD 59348.17) after $t_{0}$, with an observed magnitude of 16.76. 
The maximum $g$ brightness ($g_\mathrm{max}$) is at 13.40 days (MJD = 59343.40) after the SN explosion, with an observed magnitude of 17.12. As the $U$ light curve decays significantly more rapidly compared with $BgVri$ and there are only four detection epochs, a reliable shape of its light curve could not be determined. We present the $U$ light curve only for completeness, and it is not used in the analysis owing to these issues.

\begin{figure}
\includegraphics[scale=0.6,angle=0,width=9.1truecm]{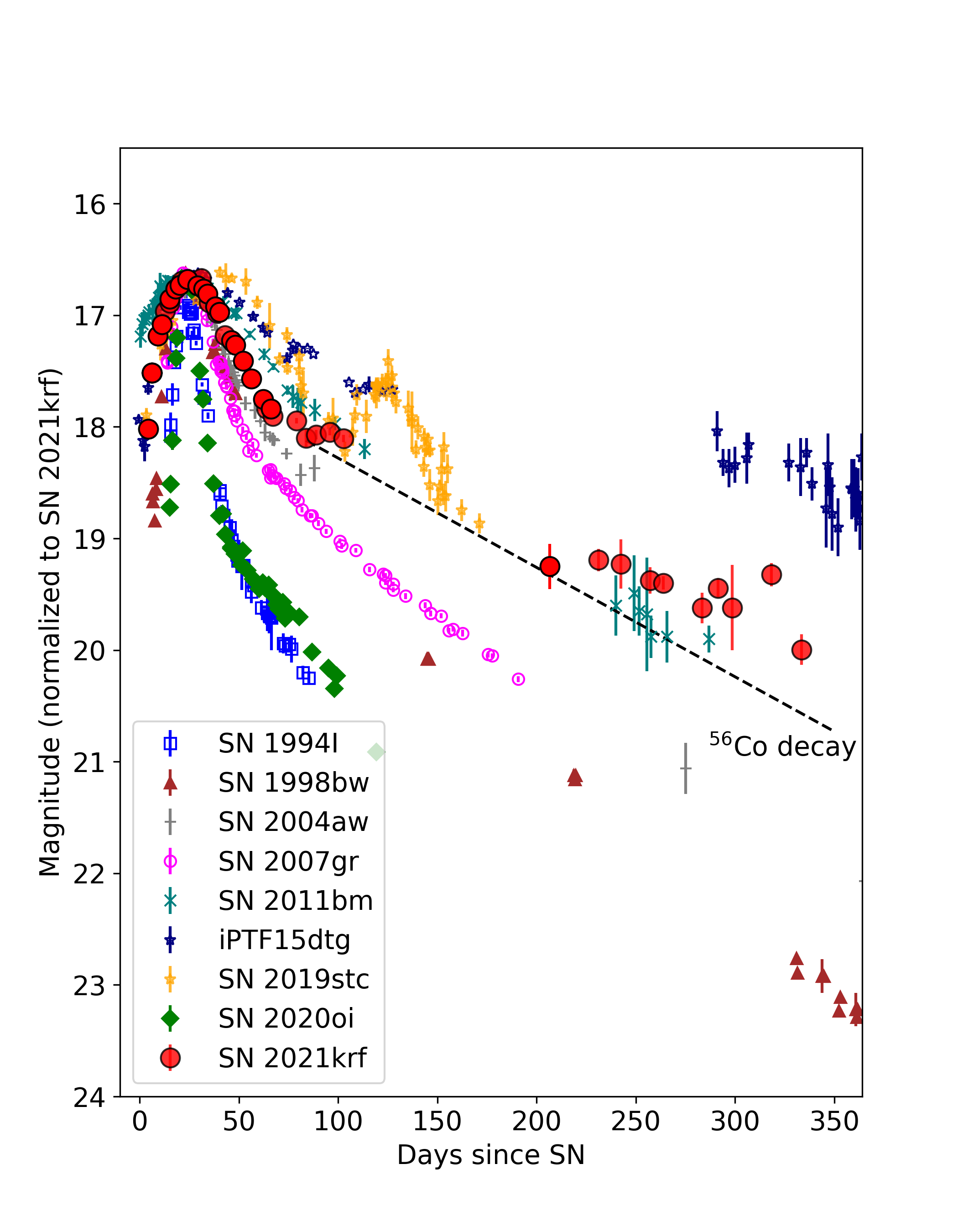}
\caption{Optical light curve of SN 2021krf in $r$ (solid red circles) compared with $r$-band light curves of other SNe~Ic.  The light curves of other SNe are scaled to match that of SN 2021krf at its peak. The $^{56}$Co decay rate is shown as a dashed line, scaled to the SN 2021krf light curve. The late-time light curve of SN 2021krf is above the $^{56}$Co decay rate, implying the existence of an additional power source.}
\label{FLCcomp1}
\end{figure}

We compare the evolution of the $r$-band light curve of SN 2021krf with those of some other SNe~Ic in Figure \ref{FLCcomp1}. The photometry of the other SNe was obtained from the Open SN catalog \citep{guillochon17}. Typically, SNe~Ic rise to peak brightness on timescales of 10 to 20\,days. The SN 2021krf light curve in $r$ peaks at $\sim 24$ days after $t_{0}$, and declines by $\sim 1.2$\,mag between days 25 and 60 (Figure \ref{FLCcomp1}). Both the rise and decline rates of SN 2021krf are slower than those of SNe~Ic such as SN 1994I \citep{richmond1996ubvri}, SN 1998bw \citep{galama98}, SN 2004aw \citep{taubenberger06}, SN 2007gr \citep{hunter09}, and SN 2020oi \citep{rho21}.

A slower declining SN~Ic light curve indicates a relatively low $E_\mathrm{k}$/$M_\mathrm{ej}$, where $E_\mathrm{k}$ is the ejecta kinetic energy and $M_\mathrm{ej}$ is the ejecta mass \citep{dessart16}. The rapidly declining light curves of SN 1994I and SN  2020oi have been ascribed to low ejecta masses \citep{filippenko95, rho21}. The light curve of SN 2007gr also declines faster compared to SN 2021krf. In contrast, SN 2011bm \citep{valenti12}, iPTF15dtg \citep{taddia19}, and SN 2019stc \citep{gomez21} have significantly slower rise and decline rates (for $t < 100$\,days), compared to SN 2021krf. This suggests that at early times ($t < 100$\,days), while SN 2021krf evolves more slowly than a typical SN~Ic, it is not as extreme as SN 2011bm, SN 2019stc, and iPTF15dtg. 

\begin{figure*} 
     \centering
     \hspace{-0.2cm}
\begin{longtable*}{cc}
\includegraphics[width=0.5\textwidth,keepaspectratio]{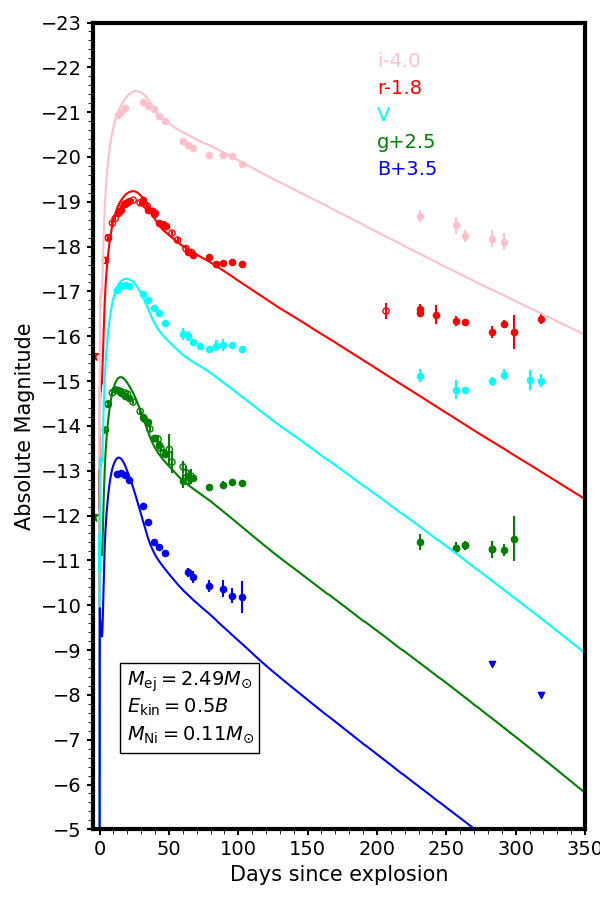} & \includegraphics[width=0.5\textwidth,keepaspectratio]{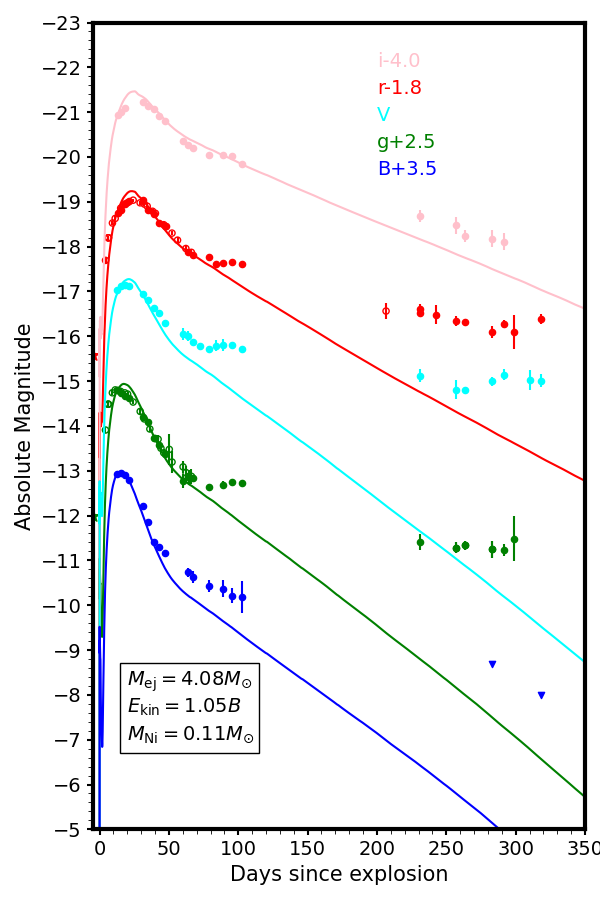} \\

(a) Progenitor: CO-3.93  & (b) Progenitor: CO-5.74 

\end{longtable*}
\caption{STELLA light-curve model fits to the observed optical light curves of SN 2021krf. Two STELLA progenitor models, (a) CO-3.93 and (b) CO-5.74, are plotted as continuous lines. Powered by Ni/Co radioactive decay, they significantly underestimate the observed fluxes in all filters after $\sim 40$ days.  In both panels, for the latest two $B$ data points, an upper limit is marked owing to high measurement uncertainties.}
\label{theory_models}
\end{figure*}

\begin{figure} 
\includegraphics[width=0.5\textwidth]{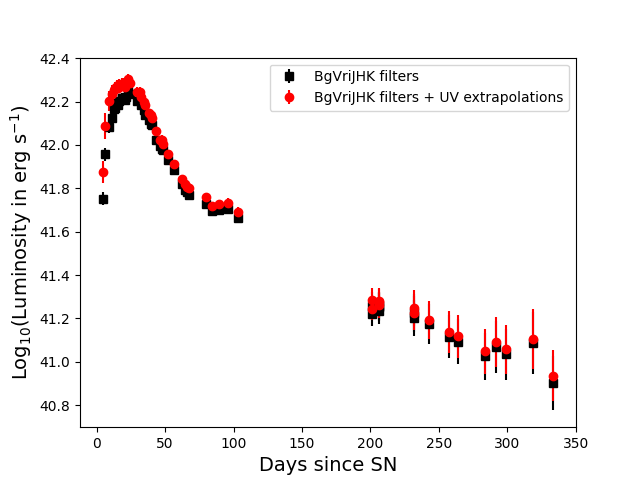}
\caption{Bolometric luminosity with $BgVriJHK$ and additional UV contributions through \textbf{SuperBOL} blackbody extrapolations.}
\label{bolometric}
\end{figure}

\begin{figure}[h]
    \centering
    \includegraphics[width=0.5\textwidth]{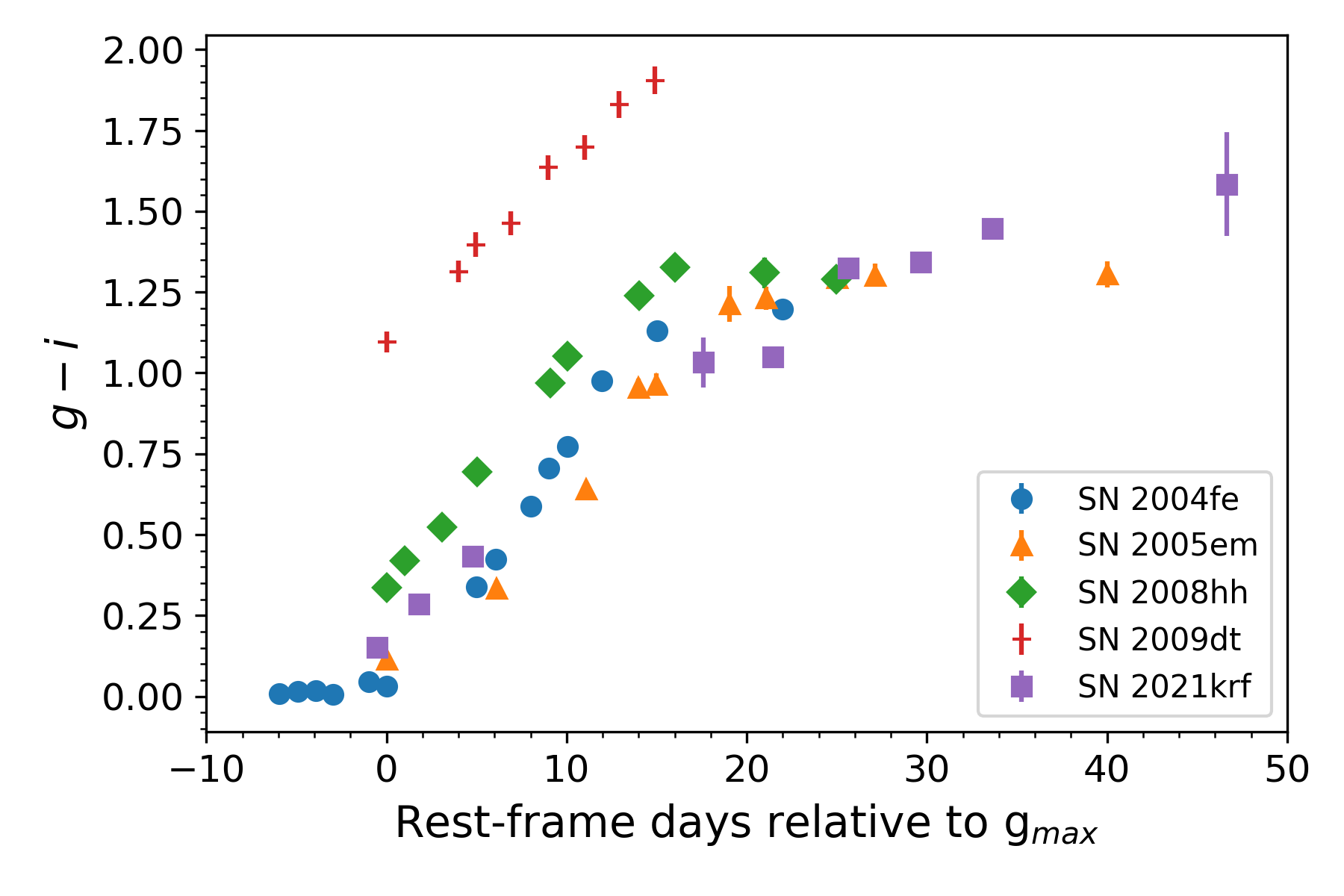}
    \caption{Evolution of $g-i$ color between SN 2021krf and the SN~Ic sample of \cite{stritzinger18a} with known low host-galaxy extinctions (SN 2004fe, SN 2005em, and SN 2008hh). 
    Corresponding upper limits of $(B-V)_{+10}$ are 0.72 mag for SN 2008hh, 1.21 mag for SN 2021krf, 1.25 mag for SN2004fe, and 1.36 mag for SN 2005em. 
    The Type Ic SN 2009dt with high extinction ($(B-V)_{+10}$ = 1.50 mag) is shown for comparison.}
    \label{extinction}
\end{figure}

A normal SN~Ic is expected to decline at least as rapidly as the $^{56}$Co radioactive decay rate \citep[see, e.g.,][]{wheeler15}, which is marked as a dashed line in Figure \ref{FLCcomp1}. At late times ($t > 200$\,days), the SN 2021krf $r$ light curve declines more slowly than the expected decay rate of $^{56}$Co (Figures \ref{Flightcurves21krf}, \ref{FLCcomp1}), which is similar to the case of SN~Ic iPTF15dtg. \cite{taddia19} showed that radioactivity along with magnetar powering is the most realistic explanation for the late-time light curve observed for iPTF15dtg. Thus, in analogy with iPTF15dtg, SN 2021krf may have additional power sources such as a central engine or CSM interaction at play during late times.

To study the pre-explosion properties of the progenitor of SN 2021krf and the explosion kinematics, we compared the observed $BgVri$ light curves from the LCO network and ZTF  with theoretical SN nucleosynthesis models obtained using the 1D multigroup radiation hydrodynamics code STELLA \citep{blinnikov98, blinnikov00, blinnikov06}. The STELLA code calculates the spectral energy distributions (SEDs) at every time step utilizing a predictor-corrector high-order implicit scheme for line emission. STELLA implicitly solves the time-dependent radiation transfer equation coupled with hydrodynamics. Individual filter light curves are obtained by convolving the corresponding filter response with the simulated SEDs.

\setcounter{table}{1}
\begin{table*} [t]
\hspace{-0.5cm}
\caption{Explosion and Progenitor Properties} 
\begin{center}
\begin{tabular}{l|cccc|ccc}
\hline \hline
                      & SNe Ic &          &            & &SN 2021krf &\\
                    &SN 2020oi    & iPTF15dtg  &SN 2007gr & SN 1994I & CO-3.93 & CO-5.74 & Arnett model$^{e}$ \\
\hline

References$^a$      &9 &8 & 1, 2, 3, 4 & 5, 6, 7 & this work & this work & this work \\

C-O star (M$_{\odot}$)$^b$  & 2.16 & ... & 1 & 2.1 & 3.93 & 5.74 & ... \\

Explosion energy $E_\mathrm{exp}$ ($10^{51}$ erg)  & 1 & ... & ... & ...  & 0.30 & 0.95 & ... \\

Kinetic energy $E_\mathrm{k}$ ($10^{51}$ erg)    & 0.6 &0.8 & 1-4 & 1 &0.50 & 1.05 & $0.73 \pm 0.23$\\

Ni mass $M_\mathrm{Ni}$ (M$_{\odot}$)$^c$   & 0.07 &0.29 & 0.076 & 0.07 & 0.11 & 0.11 & $0.118 \pm 0.007$ \\

Ejecta mass $M_\mathrm{ej}$  (M$_{\odot}$) & 0.71 &3.5 & 1.8 & 0.6 & 2.49 & 4.08 & $2.76 \pm 0.44$\\

$E_\mathrm{k}$ / $M_\mathrm{ej}$  & 0.84 & 0.23 & 0.79  & 1.67 & 0.20 & 0.25 \\

Progenitor mass (M$_{\odot}$)$^d$ & 13 & <35 & 28& 15 & 20 & 30 & ... \\

\hline
\end{tabular}
\end{center}
\hspace{0.2cm}
\renewcommand{\baselinestretch}{0.8}

{\footnotesize $^a$ (1) \cite{valenti08b}; (2) \cite{hunter09}; (3) \cite{mazzali10}; (4) \cite{crockett08};
 (5) \cite{iwamoto94}; (6) \cite{sauer06}; (7) \cite{immler02}; (8) \cite{taddia19}; (9) \cite{rho21}.}\\
{\footnotesize $^b$ C-O star mass is the progenitor mass at the pre-SN stage.}\\
{\footnotesize $^{c}$ $^{56}$Ni mixing throughout 90$\%$ of the inner ejecta assumed. }\\
{\footnotesize $^{d}$ Zero age main sequence (ZAMS) mass of the progenitor star.}\\
{\footnotesize $^{e}$ Arnett-model \citep{arnett82} fits to the photospheric-phase bolometric light curve as discussed in Section \ref{sec:4.3}}\\
\label{Tprogenitor}
\end{table*}

In general, carbon-oxygen (C-O) SN progenitors, formed through losses of their hydrogen and helium envelopes, are triggered by Fe-core collapse, creating SNe~Ic. Implementing the methodology of \cite{yoon19}, we utilize two C-O progenitor models to reproduce multicolor light curves of SN 2021krf and derive its explosion parameters. The first SN model (Model CO-3.93 in Table \ref{Tprogenitor}) adopts a helium-poor C-O progenitor mass of 3.93 M$_{\odot}$ while the initial (zero age main sequence; ZAMS) mass of the progenitor is assumed to be $\sim 20$ M$_{\odot}$. The adopted mass cut where the SN energy is injected in mass coordinates for CO-3.93 is 1.44 M$_{\odot}$, which corresponds to the outer boundary of the iron core. The second model (Model CO-5.74 in Table \ref{Tprogenitor}) assumes a helium-poor C-O progenitor mass of 5.74 M$_{\odot}$ while the initial mass of the progenitor is assumed to be $\sim 30$ M$_{\odot}$. The adopted mass cut where the SN energy is injected in mass coordinates for CO-5.74 is 1.66 M$_{\odot}$.

\begin{figure*}
\centering
\includegraphics[width=\textwidth]{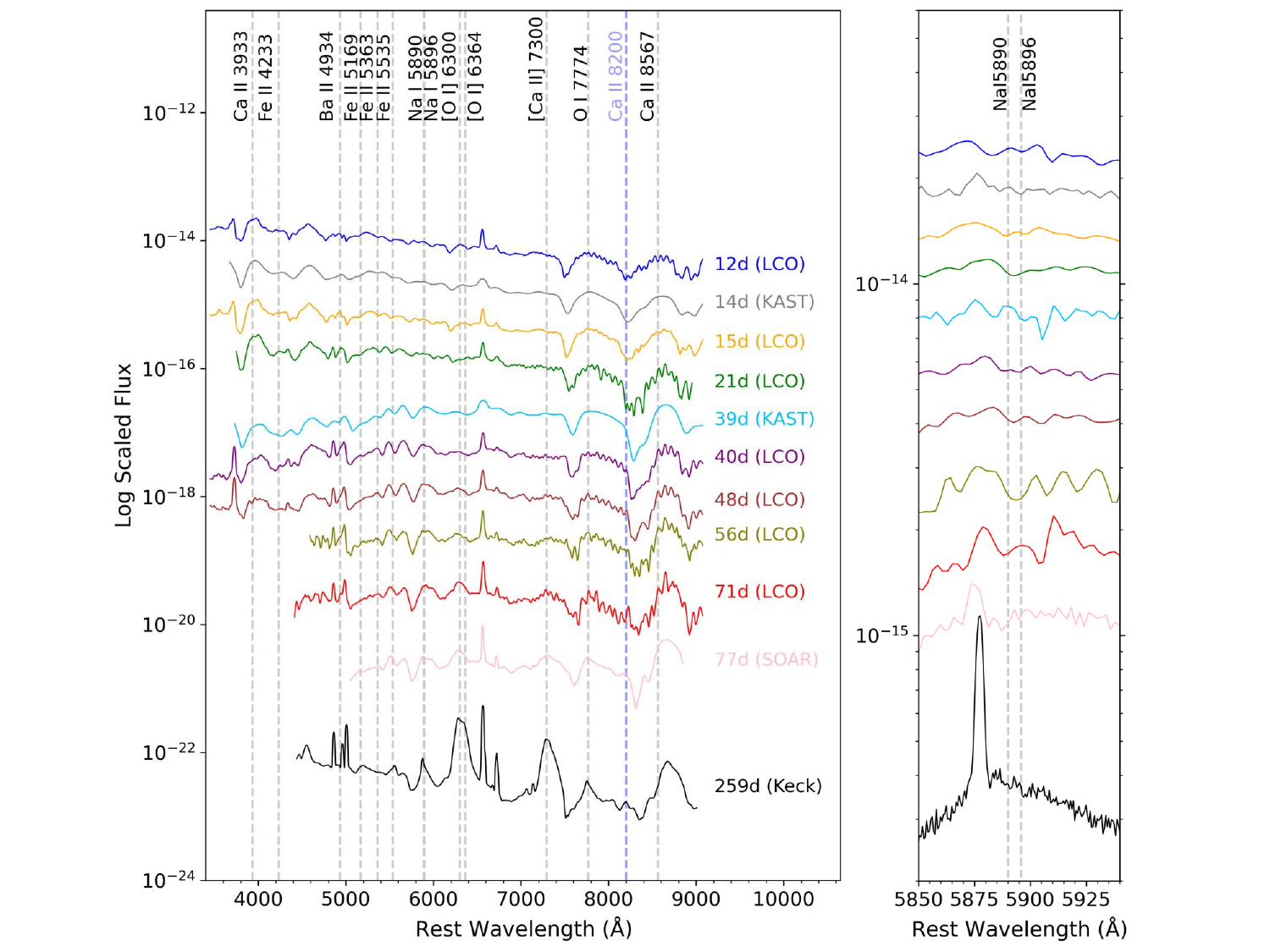}
\caption{\textit{Left:} Optical spectra of SN 2021krf obtained by the LCO network, Kast spectrograph, SOAR, and Keck-II telescopes (see Table \ref{SN21krfSpec}) after being corrected for the host galaxy's redshift ($z = 0.0135$). Fluxes for individual spectra are scaled for the purposes of display. Rest-frame wavelengths of several atomic lines of interest are marked with vertical gray dashed lines. Ca~II at 8200\,\AA\ represents the blueshifted Ca~II NIR triplet (blue dashed lines). \textit{Right:} Zoom-in view around the Na~I~D doublet wavelength range of the corresponding unbinned optical spectra at the same epochs as the left panel. No significant absorption doublet is identified; only marginal dip-like features are observed in the wavelength range}. 
\label{Foptispec21krf}
\end{figure*}

Using model CO-3.93 to fit the observed light curves, the best-fit kinetic energy ($E_\mathrm{k}$) is  $0.5 \times 10^{51}$ erg. The best-fit $^{56}$Ni ($M_\mathrm{Ni}$) and ejecta ($M_\mathrm{ej}$) masses are 0.11 M$_{\odot}$ and 2.49 M$_{\odot}$, respectively. We obtained similarly good fits with the second model CO-5.74, where the best-fit kinetic energy, $^{56}$Ni mass, and ejecta mass are $E_\mathrm{k} = 1.05 \times 10^{51}$ erg, $M_\mathrm{Ni} = 0.11$ M$_{\odot}$, and $M_\mathrm{ej} = 4.08$ M$_{\odot}$, respectively. In these models, we assume that $^{56}$Ni is uniformly mixed throughout 90$\%$ of the inner ejecta. 
\begin{figure*} [htb]
\centering

\includegraphics[width=0.70\textwidth]{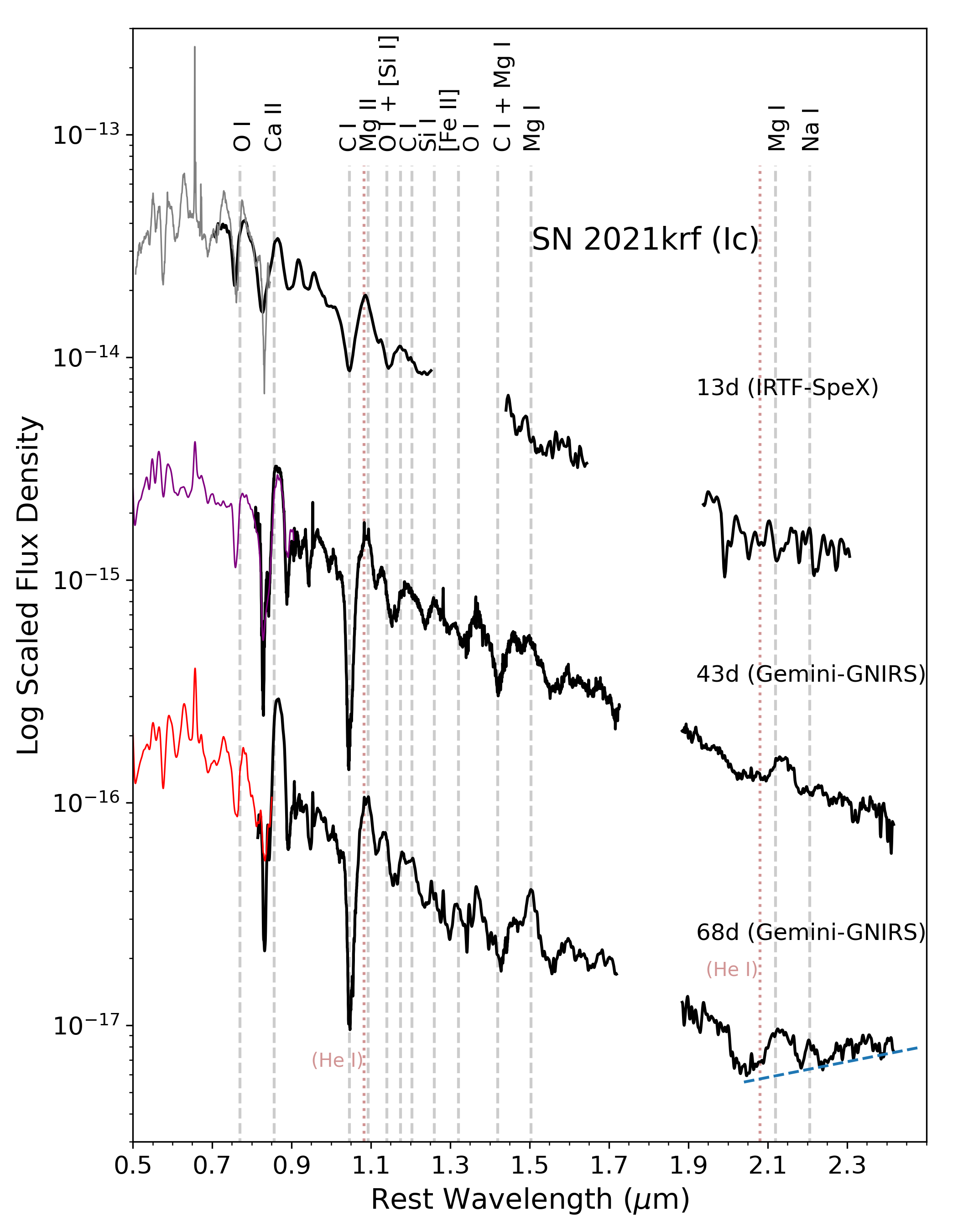}
\caption{NIR Spectra of SN 2021krf from IRTF-SpeX and Gemini-GNIRS (see Table \ref{SN21krfSpec}) after being corrected for the host galaxy's redshift ($z = 0.0135$). Optical spectra at epochs closest to the corresponding NIR epochs are included. Dashed line (blue) in the day 68 spectrum represents the approximate $K$-band continuum. Rest-frame wavelengths of several atomic lines of interest are marked with vertical gray dashed lines. Possible minor contributions from  He~I lines are marked as brown dotted lines.
}

\label{FnearIRspec21krf}
\end{figure*}

Comparisons between the observed and model light curves are presented in Figure \ref{theory_models}. Together, these models provide good fits to the observed light curves until $\sim 40$ days. Beyond  this period, the radioactive decay hydrodynamic models significantly underestimate the observed light curves. Our radiation-hydrodynamics modeling with STELLA confirms that standard radioactive decay alone cannot adequately fit the observed light curves across all bands at $t > 40$ days (Figure \ref{theory_models}).

Between days 80 and 100 after the SN explosion, the light curves in several bands ($g$, $V$, $r$, $i$) seem to show a marginal rise or bump-like feature. Similar postmaximum bumps around 80 to 150 days (for SNe~Ic) have been previously noted, such as in SN 2019stc \citep{gomez21}. They suggested that the origin of the bump feature is most likely due to a delayed circumstellar interaction with a shell ejected prior to the SN explosion. However, \cite{chugai22} argued that the second peak in SN 2019stc might have been caused by variations in the emission due to magnetar dipole field enhancement of an underlying central engine. While the secondary bump in SN 2019stc was prominent ($r = 19.54$ mag) compared to the peak magnitude ($r = 18.74$ mag), the bump-like feature in SN 2021krf (Figures \ref{Flightcurves21krf}, \ref{FLCcomp1}) is significantly fainter ($r = 18.06$ mag) compared to the earlier peak magnitude ($r = 16.66$ mag). The origin of such a bump-like feature in the light curve of SN 2021krf might have been due to additional power sources such as CSM interaction or an underlying central engine. However, with a significantly fainter bump-like feature and fewer epochs of our data at these times, we cannot make any clear inferences. We discuss the possibility of possible central-engine powering in Section \ref{sec:4.1} and CSM interaction in Section \ref{sec:4.2}.

\subsection{Bolometric Light Curve} \label{sec:3.2}

We constructed a bolometric light curve of SN 2021krf using \textbf{SuperBOL} \citep{nicholl18}, which is shown in Figure \ref{bolometric}. The input LCO and ZTF photometric magnitudes have been dereddened based on $E(B - V)$ (see Section \ref{sec:3.3}) in the direction of SN 2021krf. Photometric data from all filters (\textit{BgVri}) were interpolated to a common set of epochs and converted to flux units. We also include \textit{JHK} photometry at days 43 and 68 (Table \ref{SN21krfSpec}). Based on these fluxes, \textbf{SuperBOL} computes a quasibolometric light curve by integrating over the range of the observed filters. A blackbody function is fit to the SED, and extrapolated to include the UV contribution to the total bolometric luminosity. Including the NIR photometry greatly constrains the contribution of NIR flux at  early times ($t < 100$ days) to the total luminosity. We use the constructed bolometric light curve based on \textit{BgVriJHK} for our modeling. We discuss the evolution of Bolometric light curve in Section \ref{sec:4.3}.

\subsection{Extinction} \label{sec:3.3}
The strength of the Na~I~D $\lambda\lambda$5890, 5896 absorption doublet is indicative of the amount of dust along the line of sight. For SN 2021krf, significant absorption dips are not observed at these wavelengths (see Figure \ref{Foptispec21krf}); only marginal dip-like features are observed. We estimate the Galactic reddening toward SN 2021krf using the Galactic dust model of \cite{schlafly11}\footnote{https://irsa.ipac.caltech.edu/applications/DUST/} and obtain $E(B - V) = 0.0166 \pm 0.0004$\,mag. A correlation between extinction in the Milky Way and the equivalent width of Na~I~D lines ($EW_\mathrm{Na~I~D}$) was presented by \cite{poznanski12}.  Based on this empirical relation, the calculated value of $E(B - V)$ implies negligible absorption in the Na~I~D doublet \citep{phillips13}. This is in agreement with our observations, suggesting low extinction or a small foreground absorption in the direction of SN 2021krf.

In addition to the relation between $EW_{\mathrm{Na~I~D}}$ and host-galaxy excess for an extinction estimate, we also compare the color evolution of SN 2021krf with that of well-known SNe~Ic where the host extinction is minimal \citep[SN 2004fe, SN 2005em, SN 2008hh -- ][]{stritzinger18a}. Photometry  of SN 2004fe, SN 2005em, and SN 2008hh was obtained from \cite{stritzinger18b}. Based on an $EW_{\mathrm{Na~I~D}}$ upper limit of 1.2\,\AA\ estimated for SN 2021krf, the corresponding upper limit on the $B-V$ color at 10 days past $V$-band maximum ($(B-V)_{+10}$) is 1.22\,mag \citep[see Equation 2 of][]{stritzinger18a}. Similar upper limits on $(B-V)_{+10}$ for SN 2004fe, SN 2005em, and SN 2008hh are 0.78, 1.35, and 1.25\,mag, respectively. The $B-V$ color evolution of SN 2021krf is evidently comparable to that of other low-host-extinction SNe~Ic.

In Figure \ref{extinction}, we show the evolution of the $g-i$ color of SN 2021krf in comparison with that of SNe having low extinction and a Type Ic SN 2009dt having relatively high extinction \citep[$(B-V)_{+10} = 1.5$\,mag;][]{stritzinger18a}.
At $g_\mathrm{max}$, we see that the $(g-i)$ color observed in SN 2021krf ($\sim 0.15$\,mag) is between that of SN 2004fe ($\sim 0.01$\,mag), SN 2005em ($\sim 0.11$\,mag), and SN 2008hh ($\sim 0.31$\,mag). It is significantly different from that of SN 2009dt ($\sim 1.09$\,mag). Thus, based on both $B-V$ and $g-i$ color evolution, SN 2021krf has a host extinction comparable to that of other SNe~Ic having low host nextinctions.

\subsection{Optical and NIR Spectroscopy} \label{sec:3.4}

Figure \ref{Foptispec21krf} shows 11 optical spectra of SN 2021krf, taken between 12 and 259 days. Similarly, the NIR spectra taken between 13 and 68 days are shown in Figure \ref{FnearIRspec21krf}, along with corresponding optical data from the nearest epochs. Both optical and NIR spectra are corrected for the host-galaxy redshift with the former being expressed in standard temperature and pressure (STP) using angstroms and the latter in vacuum using microns. Rest wavelengths of atomic lines observed in the optical and NIR regimes are from the SN models by \cite{dessart12a}, synthetic spectra using SYNAPPS \citep{thomas11}, and other observed spectra of SNe~Ic \citep{gerardy02a, hunter09, drout16, jencson17, stevance17, stevance19}.

The optical and NIR spectra of SN 2021krf are dominated by atomic lines in absorption, emission, and mixed contributions in the form of P~Cygni profiles. The strongest absorption feature in the optical spectra is marked as Ca~II at 8200\,\AA\ (in Figure \ref{Foptispec21krf}), which contains contributions from the blueshifted NIR Ca~II triplet (${\lambda}$ = 8492, 8542, and 8662 \AA). The optical spectra do not reveal any clear indication of He~I lines. Several Fe~II lines appear $\sim 21$ days after the SN around 4233, 5169, 5364, and 5355\,\AA. The Na~I doublet emission first appears $\sim 40$\,days after the SN and increases in strength thereafter. Similar evolution of several other emission lines (e.g., [O~I], Ca~II, O~I/Mg~II) is also evident (Figure \ref{Foptispec21krf}). Clear [O~I] doublet line emission at 6300 and 6364\,\AA\ appears in the ejecta starting on day 48 and strengthens by day 71 at the beginning of the nebular phase. [O~I] emission is extremely strong at late times ($\sim 259$ days), indicating the existence of significant amount of oxygen ejecta as one would expect from a stripped-envelope SN.

The strongest ionic contributions in the NIR spectra are from Ca~II, O~I, C~I, Mg~II, and Si~I (Figure \ref{FnearIRspec21krf}). The Ca~II NIR triplet shows mixed contributions from absorption and emission, thus having a P~Cygni profile. We adopt a mean wavelength of 0.8567\,${\mu}$m for this triplet.  We observe significant absorption at $\sim 1.04\,\mu$m which could be blueshifted C~I at 1.0693\,$\mu$m and/or Mg~II at 1.0927\,$\mu$m. Another possible contribution to the absorption is from He~I 1.083\,$\mu$m, which is the strongest He~I transition \citep{swartz93} and thus should be sensitive to small quantities of helium \citep{wheeler93, baron96}. An in-depth analysis to estimate the relative contribution of individual ions to this feature is beyond the scope of our work.

Based on a statistically significant sample of NIR observations of stripped-envelope SNe, \cite{shahbandeh22} showed that a strong discriminator between ``He-rich'' and ``He-poor'' samples is the He~I 2.0581\,$\mu$m transition. Recent theoretical simulations by \cite{williamson21} of SNe~Ic utilizing a Monte-Carlo radiative transfer code (TARDIS) showed that for SN 1994I, even a small amount of He should produce strong optical and NIR spectral features around 5876\,\AA\ and 2.0581\,$\mu$m, respectively. They suggested that a lack of the observed absorption at 2.0581\,$\mu$m in SN 1994I is indicative of the ejecta being ``He-poor.'' The He~I absorption feature was not detected in SN 2021krf (Figure 7). Although we cannot completely rule out the presence of He in the SN ejecta owing to its possible contribution to the spectrum at 1.04\,$\mu$m, we suggest that the contribution of He-rich ejecta in the observed NIR spectrum of SN 2021krf is negligible. A similar scenario was previously suggested for SN 2007gr, where at most a very small amount of helium was reported in the NIR spectrum \citep{hunter09}. These observed line profiles along with other lines of [Si~I] at 1.129\,$\mu$m, C~I at 1.175\,$\mu$m, and Mg~I at 1.504\,$\mu$m are similar to those observed in the spectra of other SNe such as Type Ic SN 2020oi \citep{rho21}, Type IIn SN 2011dh, SPIRITS 15c \citep{jencson17}, and Type Ic SN 2007gr \citep{hunter09}.

Longward of 2\,$\mu$m, the NIR continuum flux density from SN 2021krf increases on day 68. In fact, between day 43 and 68, the slope of the spectrum above 2\,$\mu$m reversed sign (Figure \ref{FnearIRspec21krf}). This change in the NIR continuum shape suggests emission by warm dust. We discuss our spectral modeling of this continuum in Section \ref{sec:3.5}.  We note that a similar detection of warm dust was reported for the Type Ic SN 2020oi at a comparable epoch \citep{rho21}. 

\begin{figure} [htb]
\centering
\includegraphics[scale=0.5,width=8.5truecm]{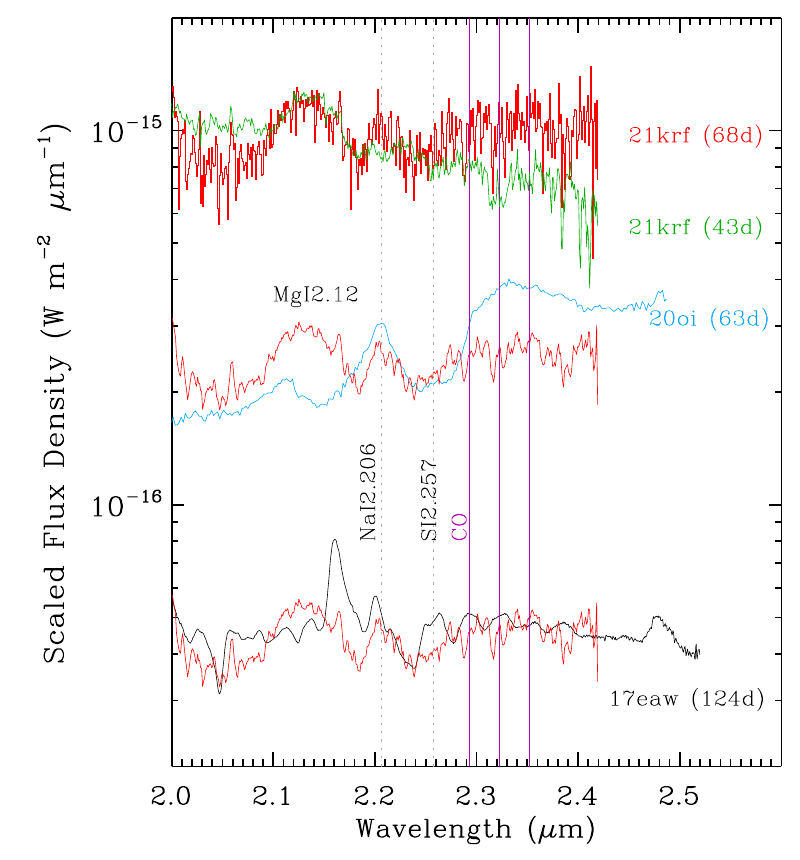}
\caption{NIR Spectra of SN 2021krf at day 68 (in red; before and after smoothing) are compared with those of SN 2021krf at day 43 (green), SN2020oi at day 63 (cyan), and SN 2017eaw at day 124 (black). The SN 2021krf spectra at day 68 show a rising continuum, but its CO bandheads (marked in purple lines at 2.293, 2.322, and 2.352\,$\mu$m) are unclear. The lack of bandheads may indicate that SN 2021krf CO is optically thick (like SN 2020oi) or has high-velocity CO. The Mg~I, Na~I, and S~I lines are marked. Type Ic SNe show a lack of S~I at 2.257\,$\mu$m compared to that of Type II-P SNe, while they enhance the Mg~I line at 2.12\,\mic.
}

\label{FnearIRspec21krfb}
\end{figure}

Figure \ref{FnearIRspec21krfb} shows a comparison of SN 2021krf spectra between 2.0 and 2.45\,\mic\ with CO bandhead (2-0, 3-1, and 4-2 at 2.293, 2.322, and 2.352\,\mic, respectively) detections from SN 2020oi (Type Ic) and SN 2017eaw (Type II-P). They are clearly evident in the spectrum of SN 2017eaw at day 124, but are not as obvious in the spectrum of SN 2021krf owing to the low SNR. Although the SNR is low, the emission at $> 2.3$\,$\mu$m on day 68 appears to rise above the continuum, as defined by the flux levels at approximately 2.10\,$\mu$m, 2.18\,$\mu$m, and 2.25\,$\mu$m. We tentatively conclude that this emission, which appears between days 43 and 68, is due to newly formed CO. Considering that the flattening or a rising continuum (evidence of dust formation) is known to occur along with CO cooling as seen in the cases of SN 1987A \citep{liu95}, SN 2017eaw \citep{rho18}, and SN 2020oi \citep{rho21}, and as expected from theoretical models \citep[e.g.,][]{sarangi13}, it would not be surprising if the first-overtone CO bands were present in SN 2021krf.

\subsection{Dust Emission in SN 2021krf} \label{sec:3.5}
\setcounter{table}{2}
\begin{table} 
    \centering
    \caption{Dust species}
    \begin{tabular}{llll}
    \hline
    Species & Grain size  & Temperature  & Dust Mass \\
            &     (${\mu}$m)            &      (K)           &   ($10^{-5}$\,M$_{\odot}$)   \\
    \hline
     C  & 0.01 & 900 $\pm$ 50 & 2.4 $\pm$ 1.1  \\
      & 0.1 & 880 $\pm$ 50 & 2.7 $\pm$ 1.2 \\
      & 1.0 & 1170 $\pm$ 90 & 0.5 $\pm$ 0.2 \\
     MgSiO$_{3}$ & 0.1 or 1 & 1020 $\pm$ 70 & 2.8 $\pm$ 1.2\\
    \hline
    \end{tabular}
    \label{dustemission_table}
\end{table} 

\begin{figure*}
\includegraphics[width=6truecm]{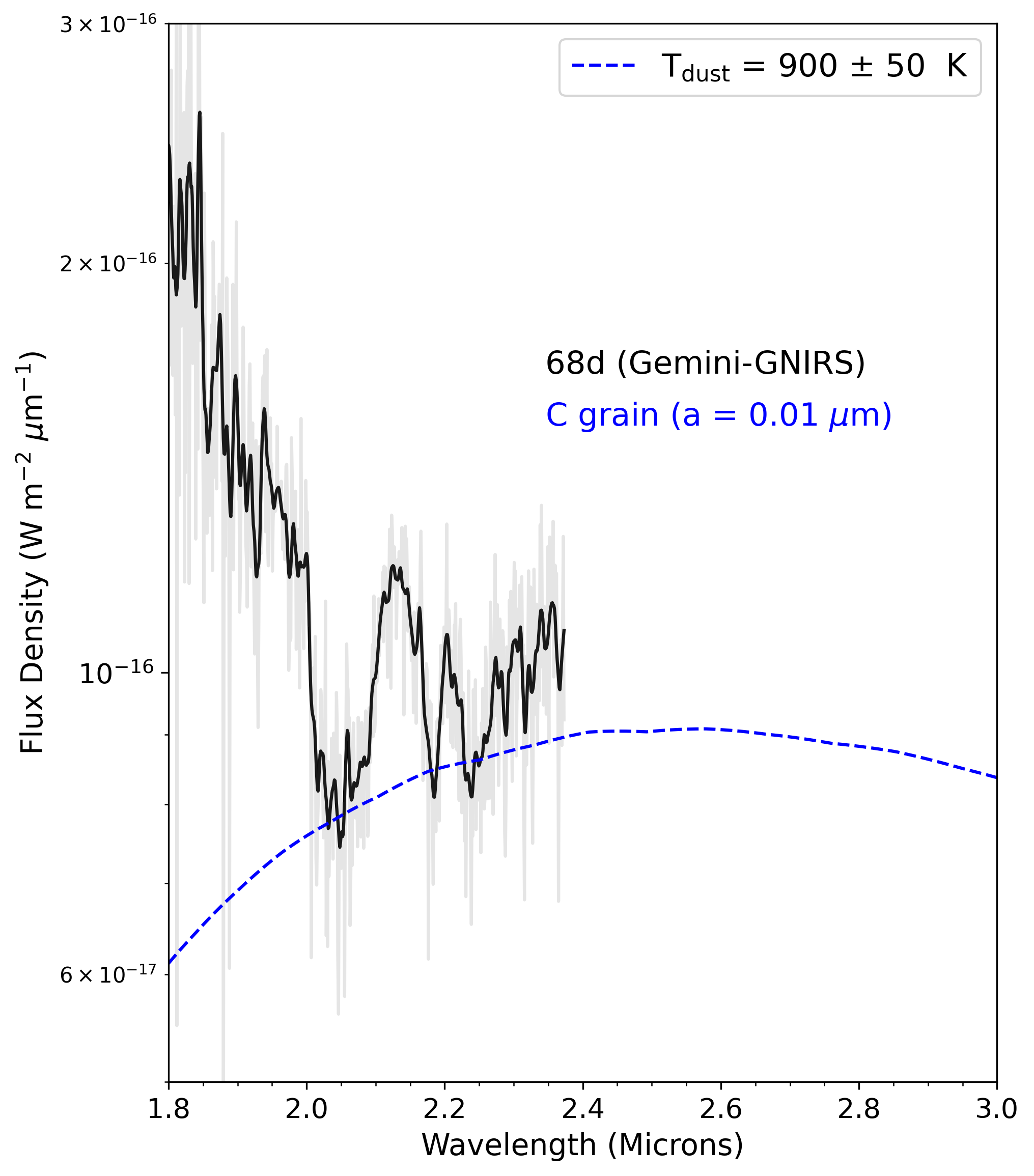}
\includegraphics[width=6truecm]{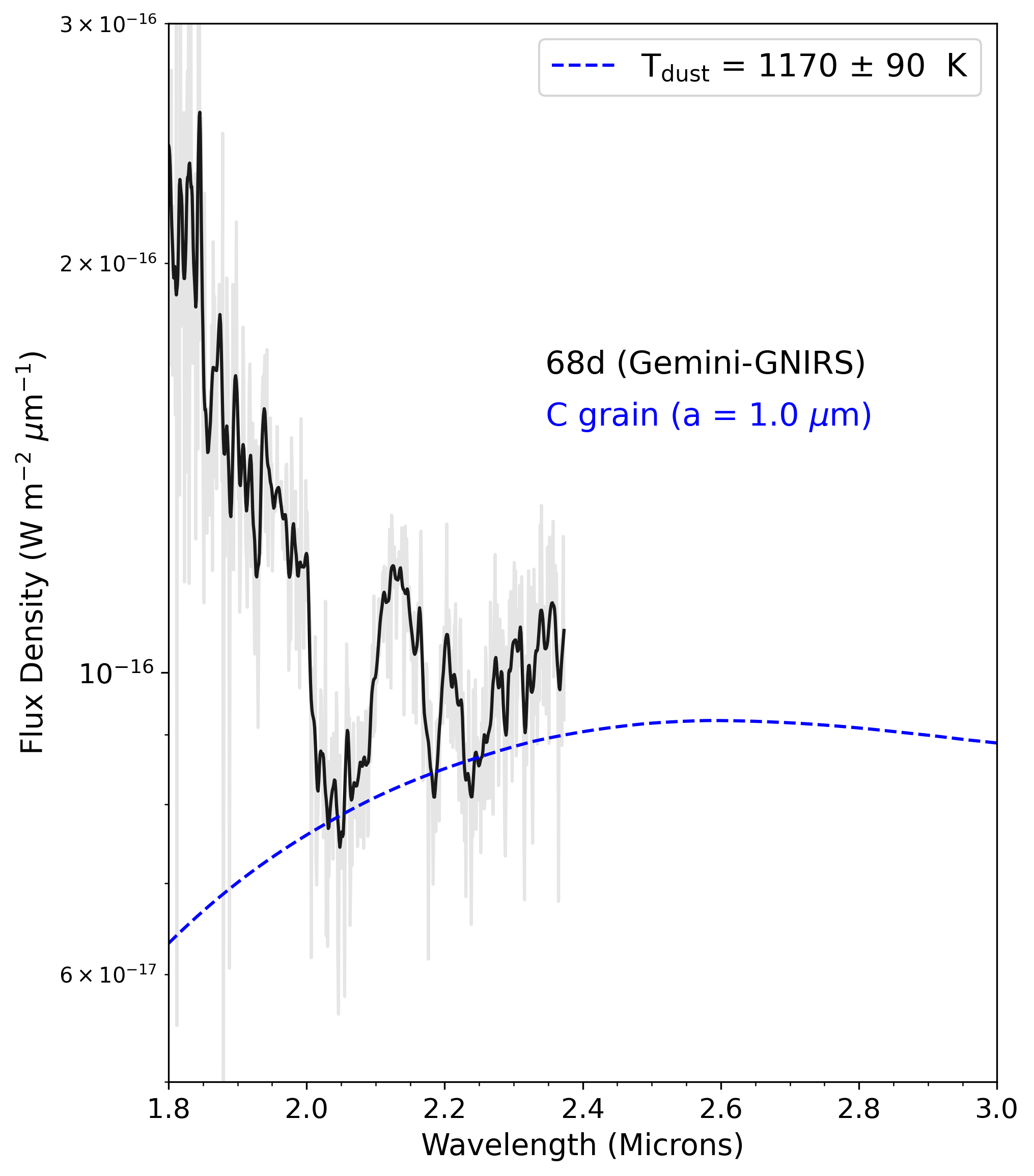}
\includegraphics[width=6truecm]{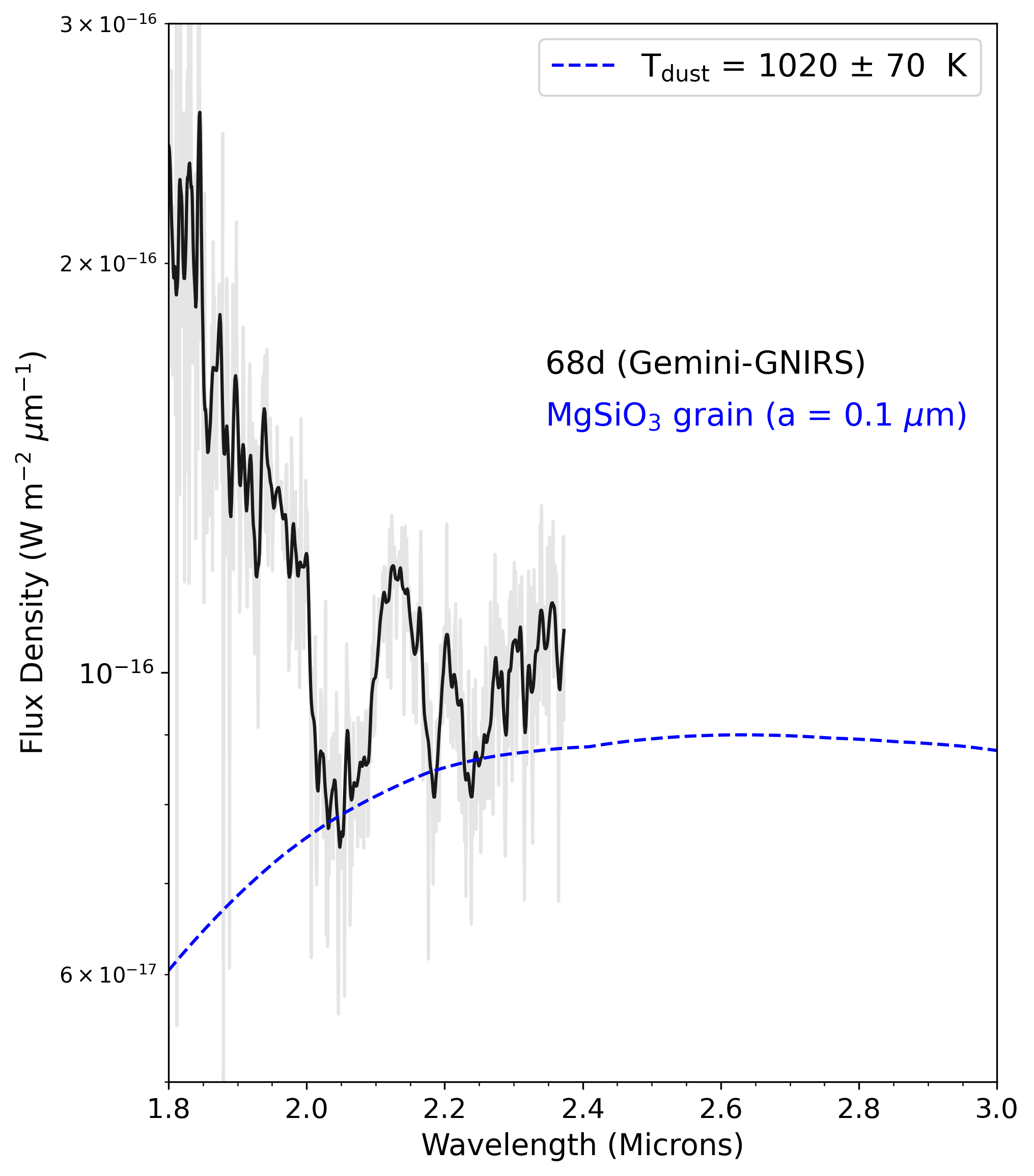}
\caption{The best-fit continuum spectral models between 2.0 and 2.3\,${\mu}$m for day 68, fit for C and MgSiO$_{3}$ dust grains of different grain sizes. The scatter in the unsmoothed data is shown in gray.}
\label{dustemission}
\end{figure*}

\begin{figure*}
\centering
\includegraphics[]{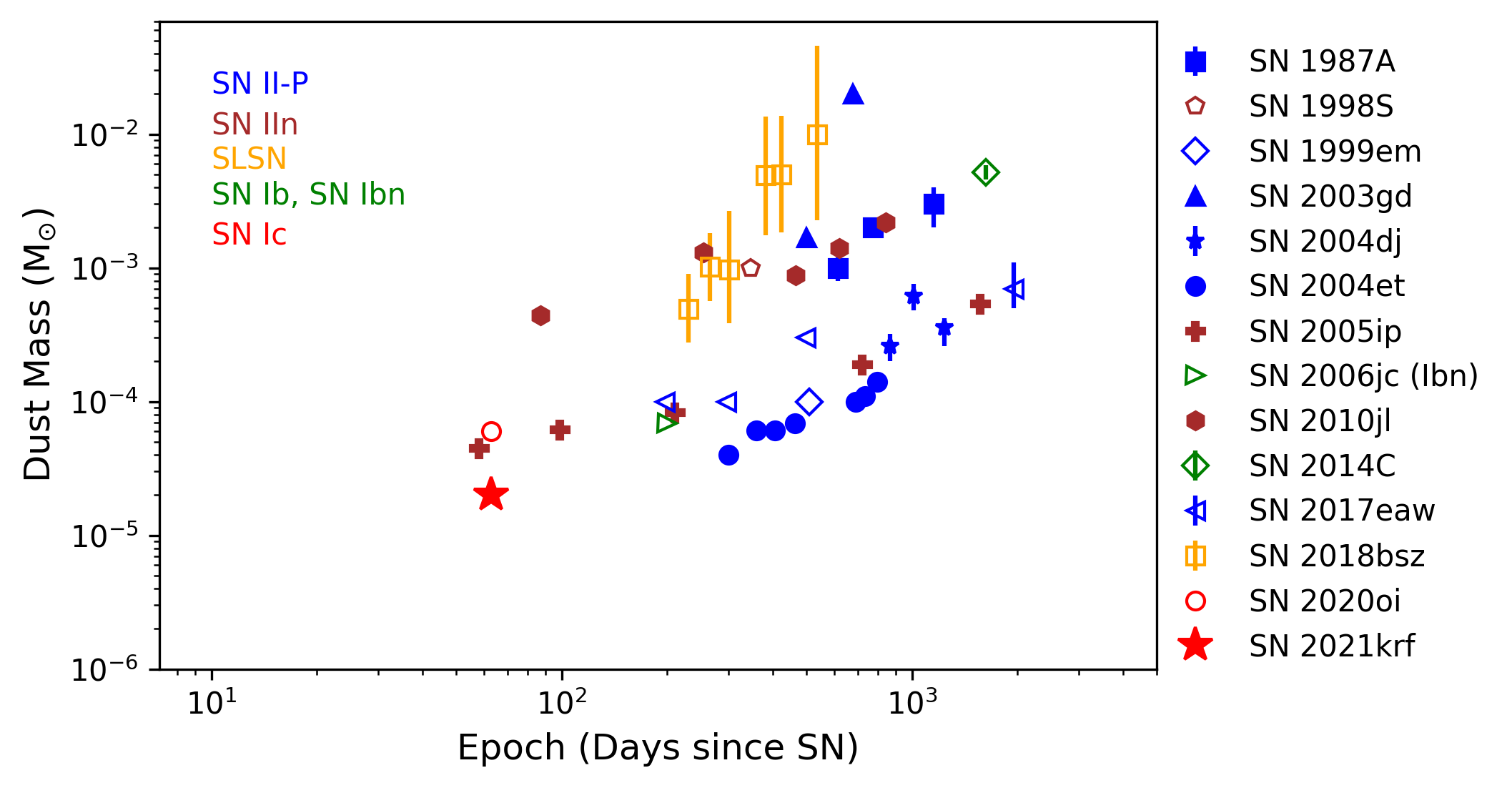}
\caption{Comparison of dust mass estimated in SN 2021krf (this work) with reported dust masses among representative cases of SN II-P (blue), SN IIn (brown), SLSN (orange), SN Ib, SN Ibn (green), and SN Ic (red). For consistency, only dust masses estimated through SED fitting have been compared. Dust masses for SN 1999em (blue diamond), SN 1998S (brown pentagon), SN 2020oi (red circle), and SN 2021krf (red star) are lower limits. The references used in the figure are as follows: SN 1987A \citep{wesson15}, SN 1998S \citep{pozzo04}, SN 1999em \citep{elmhamdi03}, SN 2003gd \citep{sugerman06}, SN 2004dj \citep{szalai11}, SN 2004et \citep{kotak09}, SN 2005ip \citep{stritzinger12}, SN 2006jc \citep{sakon09}, SN 2010jl \citep{sarangi18}, SN 2014C \citep{tinyanont19b}, SN 2017eaw \citep{tinyanont19a, shahbandeh23}, SN 2018bsz \citep{chen21}, SN 2020oi \citep{rho21}.}
\label{dust_mass_evolution}
\end{figure*}

The rising continuum probably has contributions due to CO emission (above 2.2\,\mic) as discussed above. Owing to the lack of a clear, strong detection of these CO features, we assume the features between 2 and 2.4\,\mic\ are a combination of both CO and dust continuum, similar to that of another SN~Ic, SN 2020oi \citep{rho21}.

We assume the dust grains consist of carbon, as they condense early, at temperatures of 1100--1700\,K \citep{fedkin10}. Three different grain sizes at 0.01\,$\mu$m, 0.1\,$\mu$m, and 1.0\,$\mu$m were considered. We also considered alternate silicate grains such as MgSiO$_{3}$, as they condense at temperatures of 1040--1360\,K \citep{speck11}. 
We have fitted the dust continuum (in the range 2.0--2.3\,\mic) in SN 2021krf at 68 days (Figure \ref{FnearIRspec21krf}) with a modified blackbody model, which is the Planck function, $B_{\nu}(T)$, multiplied by the absorption efficiency, $Q_{\rm abs}$. The continuum we use to fit is similar to the analysis in SN 2020oi \citep{rho21}, where we used the three portions of the continuum: 2.01--2.08\,\mic, 2.155--2.17\,\mic, and 2.255--2.285\,\mic. The portions of the continuum exclude the wavelength ranges of CO features.

The best continuum-fitting results are shown in Figure \ref{dustemission}. Optical constants of the grain species used in the calculation of $Q_{\rm abs}$ are the same as described by \cite{rho18} and references therein. The best-fit temperatures and dust masses are presented in Table \ref{dustemission_table}. Depending on these results and assumed grain species and sizes, we estimate a dust mass between $\sim$ $5 \times 10^{-6}$ and $2 \times 10^{-5}$\,M$_{\odot}$ and a dust temperature range of $\sim$ 900--1200\,K. Additional cooler dust emitting at wavelengths greater than  2.5\,\mic\ cannot be adequately constrained from our data. Thus, our derived dust mass is a lower-limit estimate to the total dust at this phase. We discuss the possible origin of this dust emission in Section \ref{sec:4.4}. 

In Figure \ref{dust_mass_evolution}, we compare the evolution of dust masses observed across multiple epochs in SNe~Ib/c, SNe~IIn, SNe~II-P, and SLSNe with the amount of dust estimated in SN 2021krf.  The dust mass in SN 2021krf is similar to that seen at similar epochs in another Type Ic SN 2020oi. The rising 2.0--2.5\,\mic\ continuum of SN 2020oi at day 63 was attributed by \cite{rho21} to freshly formed dust in the Si-S layers at a temperature of $\sim 810$\,K and a mass of $6 \times 10^{-5}$\,M$_{\odot}$. Dust signatures seen in SNe~Ic are rare; SN 2021krf is only the second case of an SN~Ic showing a rising NIR dust continuum at an early epoch. At later epochs ($t > 1.6$\,yr), \cite{kankare14} noted the presence of dust at MIR wavelengths through \textit{Spitzer} observations of SN 2005at; however, the corresponding dust mass was not estimated in their work. The amount of dust in SN 2021krf is also significantly smaller than that observed in SN 2018bsz, an SLSN \citep{chen21, pursiainen22}. Rather, it is more similar to the amounts of dust estimated in other SNe (Types Ic, Ib, IIn, II-P) at their earliest epochs (Figure \ref{dust_mass_evolution}).\\

\section{Discussion} \label{sec:4}
Spectral model fits of the optical spectra in the photospheric phase between days 12 and 71 are presented in Section \ref{sec:4.1}, and velocity profiles of the observed line features in Section \ref{sec:4.2}. We model the bolometric light curve to identify potential sources of the observed late-time emission in Section \ref{sec:4.3}. We consider multiple scenarios to explain the origin of dust emission in Section \ref{sec:4.4}.  

\subsection{Optical Spectral Modeling: Photospheric Phase} \label{sec:4.1}
\begin{figure*}
\centering
\includegraphics[width=8.5cm]{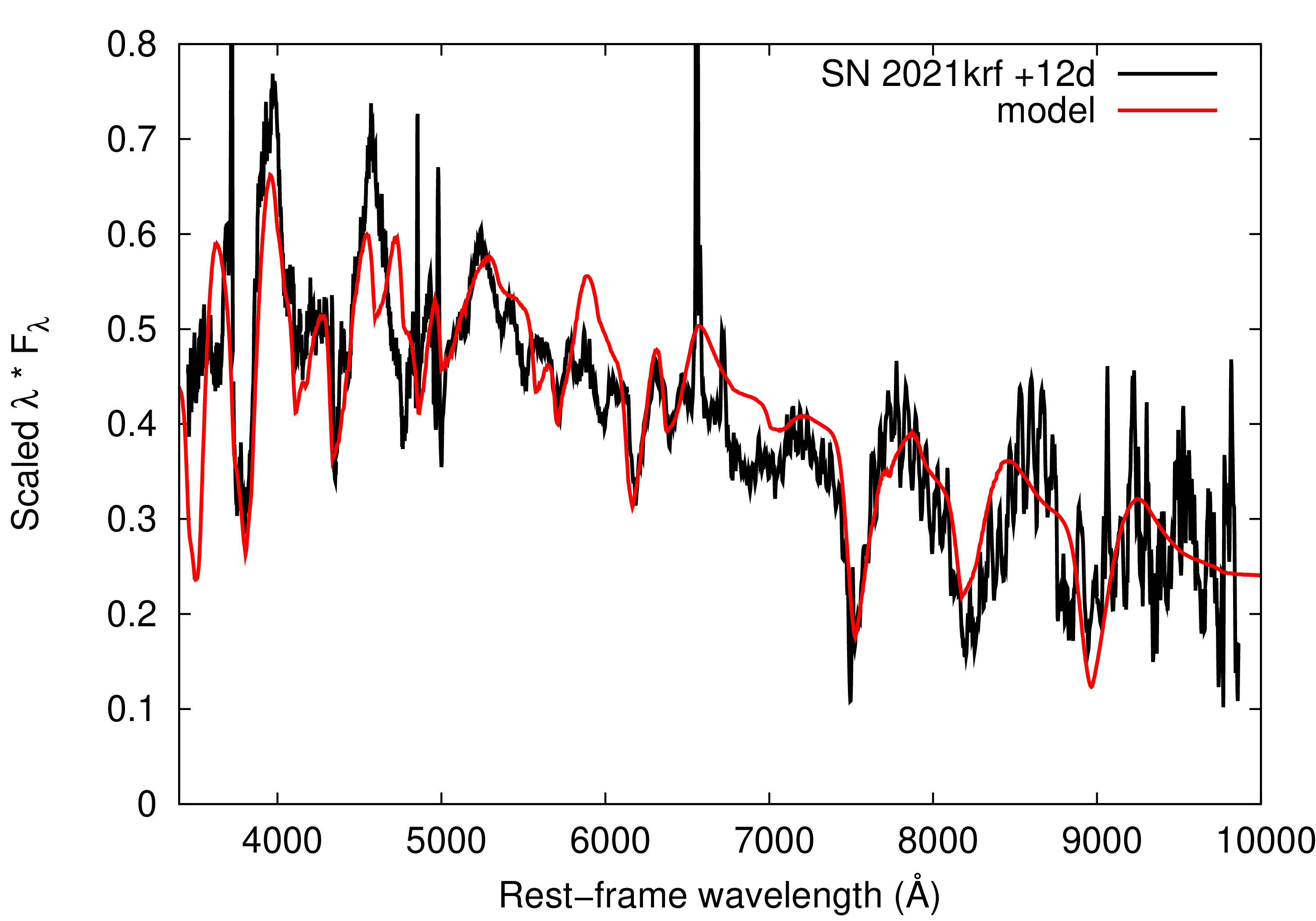}
\includegraphics[width=8.5cm]{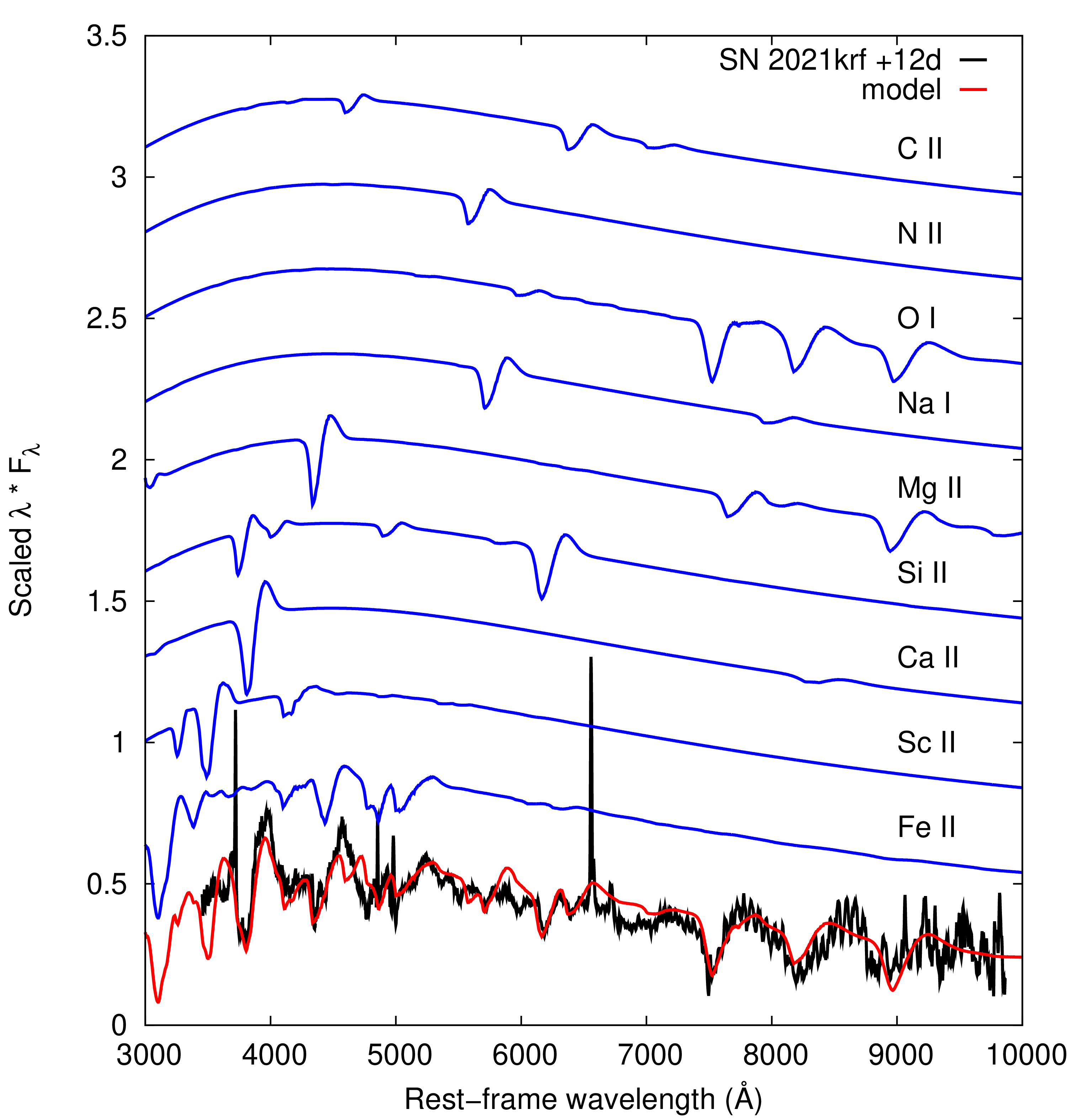}
\caption{{\it Left:} The observed spectrum of SN~2021krf (black line) at day $12$ plotted together with its best-fit model spectra obtained with {\tt SYN++} (red line).  On the ordinate $\lambda\,F_\lambda$ is plotted. {\it Right:} Single-ion contributions (blue lines) to the overall model spectrum (black line). The model spectra are shifted vertically from each other for better visibility.}
\label{fig:m1}
\end{figure*}

\begin{figure}
\centering
\includegraphics[width=8.5cm]{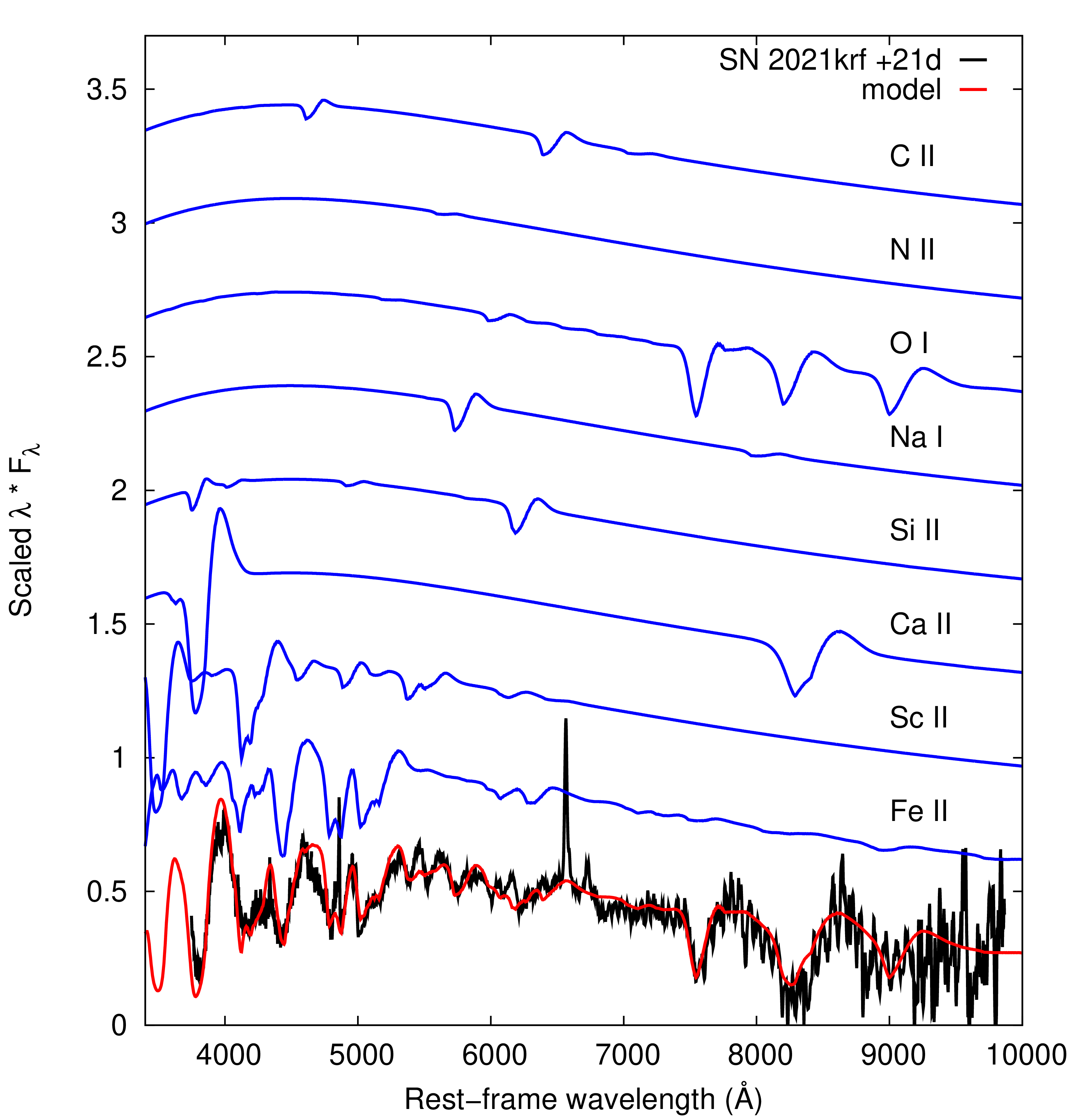}
\includegraphics[width=8.5cm]{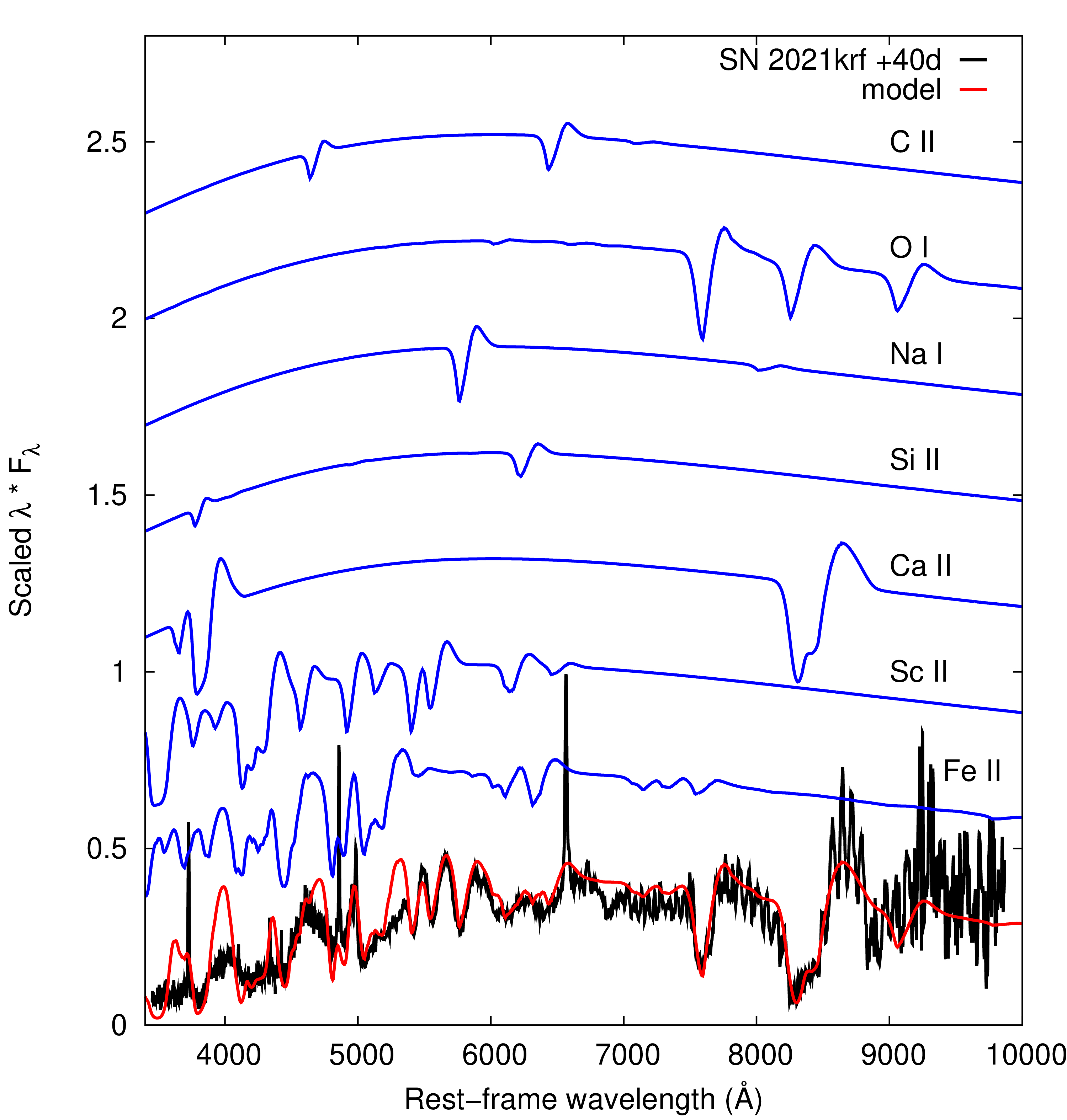}
\caption{The observed and modeled spectra of SN~2021krf at day 21 (upper panel) and day 40 (lower panel) phases. The color coding is the same as in Figure \ref{fig:m1}. }
\label{fig:8_27}
\end{figure}

\begin{figure} 
\centering
\includegraphics[width=0.5\textwidth]{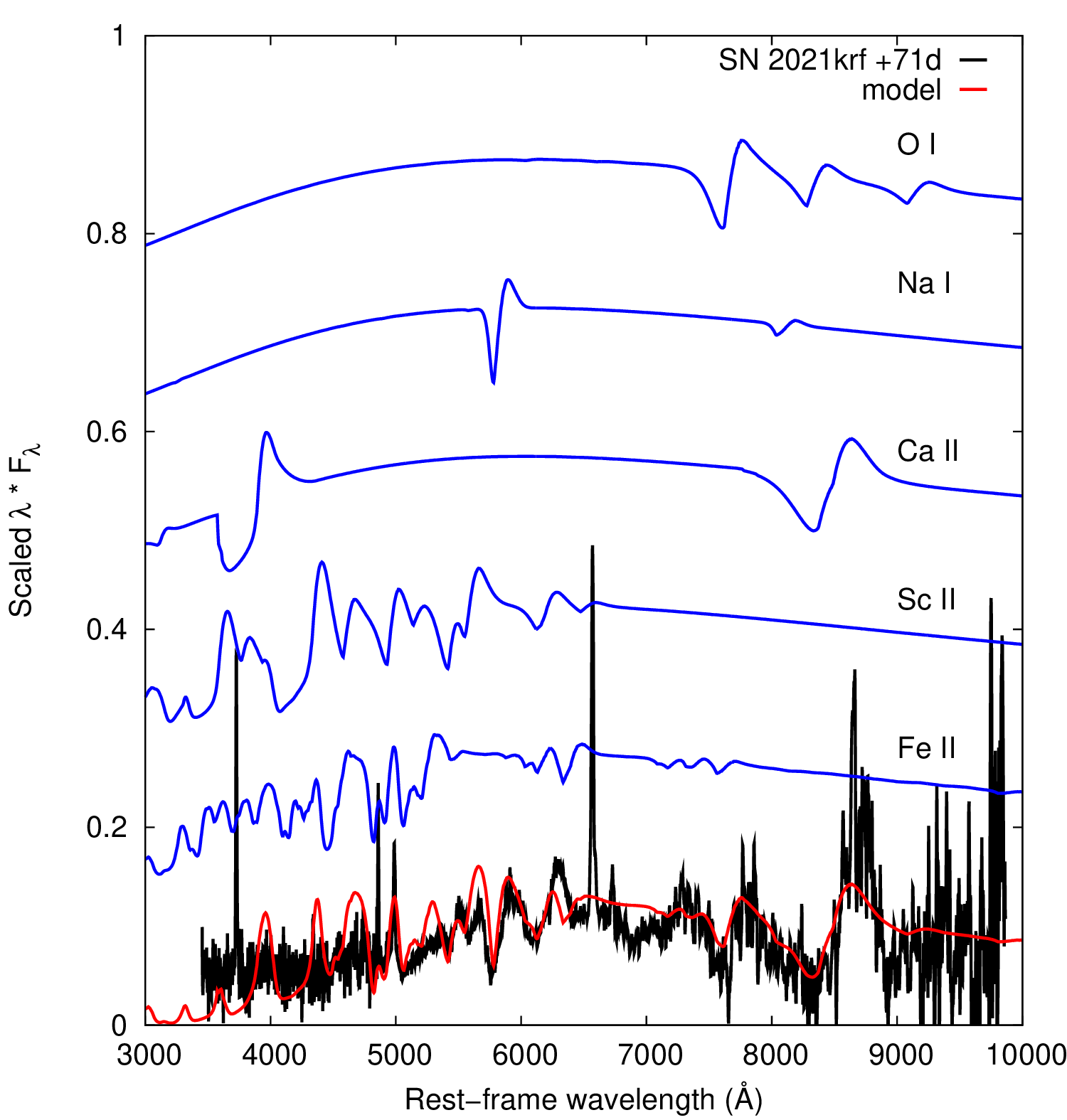}
\caption{The observed and modeled spectra of SN~2021krf at day 71 with the same coloring scheme adopted in Figure \ref{fig:m1}. Single-ion contributions to the overall model spectrum are shown.}
\label{fig:57}
\end{figure}

The photospheric phase optical spectra of SN~2021krf contain some blended P~Cygni lines.  A spectral synthesis code is required to identify these features and to examine the chemical and velocity evolution of the object. LCO spectra were modeled at days 12, 21, 40, and 71. We present the best-fit spectral model at these epochs in Figures \ref{fig:m1}, \ref{fig:8_27}, and \ref{fig:57}.

We utilized the {\tt SYN++} \citep{thomas11} code, an improved version of the original {\tt SYNOW} code \citep{hatano99}, to model the available photospheric optical spectra of SN~2021krf. This code uses some global parameters: $a_0$, a constant normalization parameter to scale overall model flux; $v_{\rm phot}$, the photospheric velocity; and $T_{\rm phot}$, the photospheric temperature. Other parameters characterize the features of different ions: $\log \tau$, the optical depth for the reference line of each ion; $v_{\rm min}$, the inner velocity of the line-forming region; $v_{\rm max}$, the outer velocity of the line-forming region; $\sigma$, the scale height of the optical depth in the line-forming region in km\,s$^{-1}$; and $T_{\rm exc}$, the Boltzmann excitation temperature of each element assuming  local thermodynamic equilibrium (LTE). All spectra were corrected for redshift and Milky Way extinction before the fitting. The calculated {\tt SYN++} model parameters can be found in Table \ref{tab:syn++}.

\begin{table}
\begin{center}
\caption{{\tt SYN++} model parameters}
\label{tab:syn++}
\begin{tabular}{lcccc}
\hline
\multicolumn{5}{c}{Global parameters} \\
\hline
Parameter & 12\,d & 21\,d & 40\,d & 71\,d \\
\hline
$v_{\rm phot}$ (km\,s$^{-1}$) & 10,000 & 9000 & 7000 & 6000 \\
$T_{\rm phot}$ (K) & 8000 & 8000 & 6000 & 6000 \\
\hline
\multicolumn{5}{c}{log optical depths} \\
\hline
Ion & 12\,d & 21\,d & 40\,d & 71\,d \\
\hline
C II & -1.5 & -2.0 & -3.9 & -- \\
N II & -1.7 & -3.0 & -- & -- \\
O I & 0.9 & 0.9 & -0.2 & -0.5 \\
Na I & 0.3 & 0.0 & -1.0 & -1.0 \\
Si II & 0.5 & 0.0 & -1.5 & -- \\
Mg II & 0.6 & -- & -- & -- \\
Ca II & 0.5 & 1.5 & 2.0 & 1.5 \\
Sc II & -0.8 & 0.2 & -0.7 & -0.3 \\
Fe II & 0.4 & 1.2 & 0.2 & -0.3 \\
\hline
\multicolumn{5}{c}{Feature widths in units of 1000\,km\,s$^{-1}$} \\
\hline
C II &  1.0 & 1.0 & 1.0 & -- \\
N II & 1.0 & 1.0 & -- & -- \\
O I & 1.0 & 1.0 & 1.0 & 3.0 \\
Na I & 1.0 & 1.0 & 1.0 & 1.0 \\
Si II & 1.0 & 1.0 & 1.0 & -- \\
Mg II & 1.0 & -- & -- & -- \\
Ca II & 2.0 & 2.0 & 1.0 & 5.0 \\
Sc II & 1.0 & 1.0 & 1.0 & 1.0 \\
Fe II & 1.0 & 1.0 & 1.0 & 1.0 \\
\hline
\multicolumn{5}{c}{Excitation temperatures in units of 1000\,K} \\
\hline
C II &  8.0 & 8.0 & 6.0 & -- \\
N II & 8.0 & 8.0 & -- & -- \\
O I & 8.0 & 8.0 & 6.0 & 6.0 \\
Na I & 8.0 & 8.0 & 6.0 & 6.0 \\
Si II & 8.0 & 8.0 & 6.0 & -- \\
Mg II & 8.0 & -- & -- & -- \\
Ca II & 8.0 & 8.0 & 6.0 & 6.0 \\
Sc II & 8.0 & 8.0 & 7.0 & 6.0 \\
Fe II & 8.0 & 8.0 & 6.0 & 6.0 \\ 
\hline
\hline
\end{tabular}
\tablecomments{For all models, $v_{\rm min} = v_{\rm phot}$ and $v_{\rm max} = 30,000$\,km\,s$^{-1}$ was assumed.}
\end{center}
\end{table}

The best-fit model of the first spectrum, taken at day $12$, shows a photospheric velocity of 10,000\,km\,s$^{-1}$, a typical value for SNe~Ic at early, pre-maximum phases. The temperature at the photosphere is 8000\,K, which is consistent with the spectrum being dominated by lines of neutral and singly ionized elements: C~II, N~II, O~I, Na~I, Mg~II, Si~II, Ca~II, Sc~II, and Fe~II. All features identified with {\tt SYN++} are consistent with being photospheric; in fact, no detached or high-velocity features were found at any epoch.

\begin{figure}
\centering
\includegraphics[width=8cm]{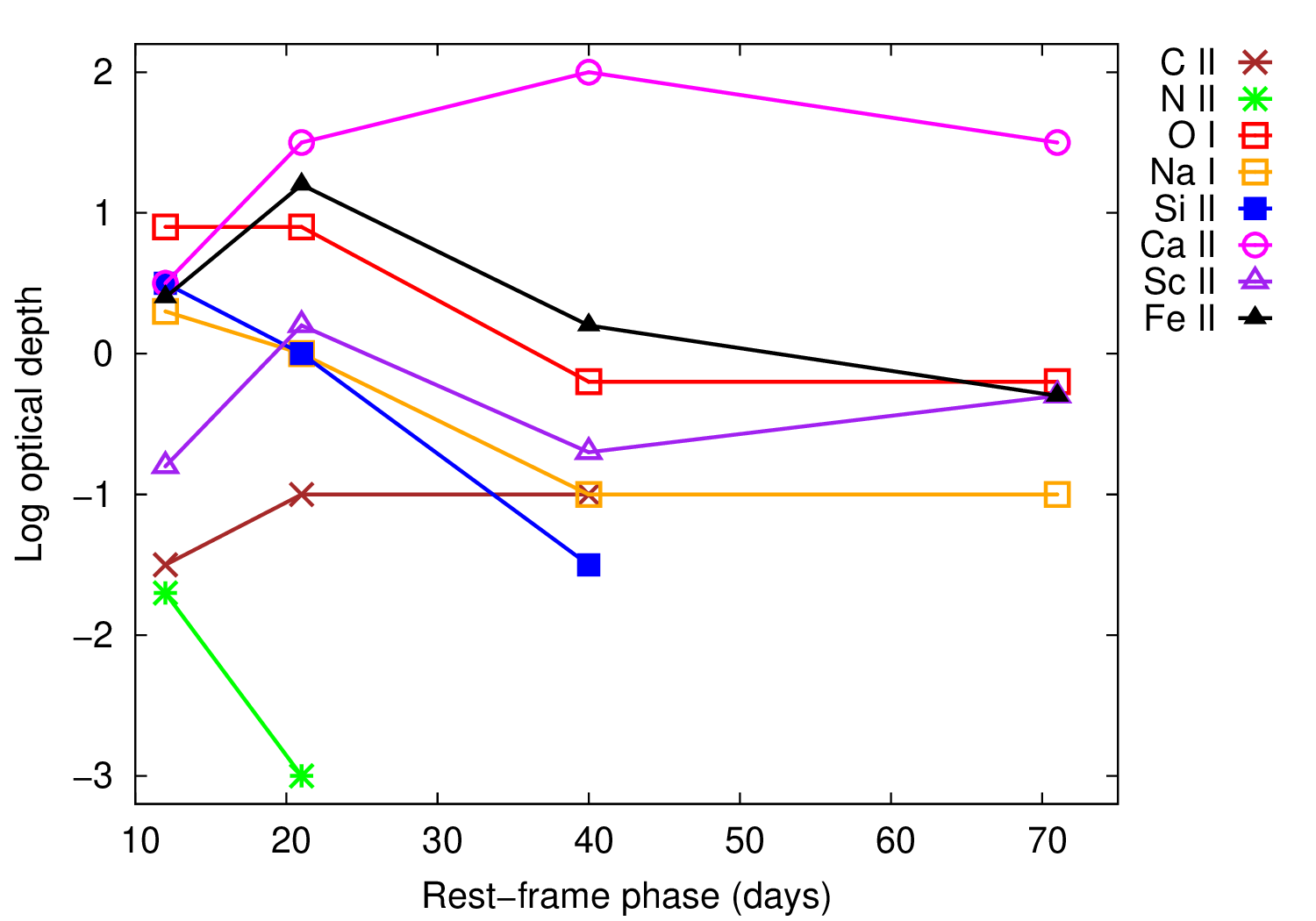}
\caption{The optical-depth evolution of the identified elements as a function of the rest-frame phase relative to the epoch of explosion.}
\label{fig:phase_vs_tau}
\end{figure}

The observed spectrum and best-fit model spectrum associated with the first epoch at day 12 are shown in the left panel of Figure \ref{fig:m1}. In the right panel, contributions of individual ions to the overall model spectrum are shown. The best-fit model of the second-epoch spectrum, which was taken at day 21, contains the same ions as the pre-maximum model, but with slightly different optical depths (see Table \ref{tab:syn++}, the top panel of Figure \ref{fig:8_27}, and Figure \ref{fig:phase_vs_tau}). The photospheric velocity and temperature did not change significantly by this epoch. Also, the Mg~II $\lambda 4481$ line was blended with the strengthening Fe~II $\lambda 4549$ feature; thus, it could not be constrained and was omitted from the model.

Including the Sc~II features improved the fitting between 4000 and 6000\,\AA, even though Sc is not commonly identified in SN~Ic spectra. Recently, \cite{gutierrez22} also found Sc~II in spectra of SN 2020wnt, a peculiar hydrogen-poor Type I SLSN. Since SLSN-I spectra are somewhat similar to those of SNe~Ic, the presence of Sc~II is not unexpected. According to \cite{hatano99}, Sc~II can reach $\log \tau \gtrsim -1$ in C/O-rich SN ejecta when the temperature is below 8000\,K. This is consistent with our modeling results (Table \ref{tab:syn++}).

The bottom panel of Figure \ref{fig:8_27} displays the modeling of the third spectrum observed at day 40. By this phase the expansion velocity of the photosphere has receded to $\sim 7000$\,km\,s$^{-1}$, and the photospheric temperature has declined from 8000\,K to 7000\,K. Because of the latter decrease, the N~II lines have disappeared from the spectrum, and the O~I, Na~I, Si~II, Sc~II, Fe~II, 
and C~II lines weakened. Only one ion, Ca~II, increased in both emission and absorption, compared to the previous epochs, as can also be seen in the evolution of optical depths of individual ions (Figure \ref{fig:phase_vs_tau}). 

The spectrum taken at day 71 is plotted along with its best-fit model in Figure \ref{fig:57}, together with the contributions of individual ions to the overall model spectrum. By this epoch, $v_{\rm phot}$ has decreased to 6000\,km\,s$^{-1}$ and $T_{\rm phot}$ to 6000\,K. The identified features are due to O~I, Na~I, Ca~II, Sc~II, and Fe~II.  The strengths of these lines have not changed significantly since the previous epoch, except for the Ca~II NIR triplet, which has further strengthened. Note that some forbidden transitions, including the [O~I] $\lambda \lambda$6300, 6364 doublet, have started to strengthen, marking the beginning of the nebular phase. 

To summarize, based on our optical spectral fitting of the photospheric phase of SN 2021krf, significant temporal changes occurred between days 12 and 71, mainly in the strong absorption due to O~I and Ca~II ions. 
The feature at $\sim5800$\,\AA\ could also be modeled by the He~I $\lambda5876$ line instead of the Na~I $\lambda\lambda 5890$, 5896 doublet. However, the optical He lines are highly blended with many other nearby lines, whereas the NIR He lines (e.g., 2.059\,\mic) are more isolated. When we examine the He absorption feature at 2.058\,\mic, there is no clear evidence of this line in SN 2021krf (see Figure~\ref{FnearIRspec21krf} and Section \ref{sec:3.4}). The evidence for He is even weaker than that discussed in the Type Ic SN 2020oi \citep{rho21}. Therefore, we identify SN 2021krf as a Type Ic SN.

\subsection{Velocity Profiles} \label{sec:4.2}

Doppler shifted line-velocity measurements based on the {\tt SYN++} model fits to the observed optical spectra are presented in Table~\ref{tab:vel}. Based on the spectral models, an uncertainty of 500\,km\,s$^{-1}$ is estimated for each individual ion line velocity. As all optical spectra have already been corrected for the host-galaxy redshift, the Doppler shift of individual lines is entirely due to kinematics in the SN. Note that these ion velocities are different from the photospheric velocity given by the {\tt SYN++} models, as they are calculated simply from the Doppler shift of the absorption minima of the strong features in the spectra, assuming that a particular feature is entirely due to a given ion. Hereafter, we refer to it as ion velocities. However, the changes in the observed spectral features could be due to temporal changes from one atomic line or a blending contribution from several atomic lines. As we have both optical and NIR spectral observations, we use the strong absorption feature produced by the Ca~II NIR triplet, observed in both the optical and NIR spectra to independently estimate its velocity.

\begin{table}
    \caption{Doppler shift velocities (\kms)}
    \begin{center}
    \begin{tabular}{llllll}
    \hline
    Age & Ca II & Fe II & Si II &  O I & Ca II \\
    (d)    & 3951\,\AA  & 5169\,\AA & 6355\,\AA & 7774\,\AA & 8566\,\AA \\
    \hline
     12 & 11500 & 9800 & 7800 & 9800 & 11200 \\
     15 & 11400 & 9600 & 7100 & 9300 & 11100 \\
     21 & 10900 & 8700 & 5500 & 7700 & 9900 \\
     40 & 11300 & 6700 & --   & 6800 & 9100 \\
     48 & 10200 & 6800 & --   & 6800 & 9000 \\
     56 & 9800  & 6800 & --   & 6600 & 9000 \\
     71 & --    & 5600 & --   & 6600 & 8800 \\
     77 & --    & --   & --   & 6500 & 8800 \\
    \hline
    \end{tabular}
    \end{center}
    \tablecomments{Doppler shift velocities of the absorption minima for several ions between 12 and 77 days. For all spectral profiles from SYN++ modeling, a constant uncertainty of 500\,km\,s$^{-1}$ was estimated. All values are quoted with appropriate significant digits.}
    \label{tab:vel}
\end{table}

\begin{figure*}
\centering
\begin{tabular}{cccc}
\includegraphics[scale=0.55,keepaspectratio]{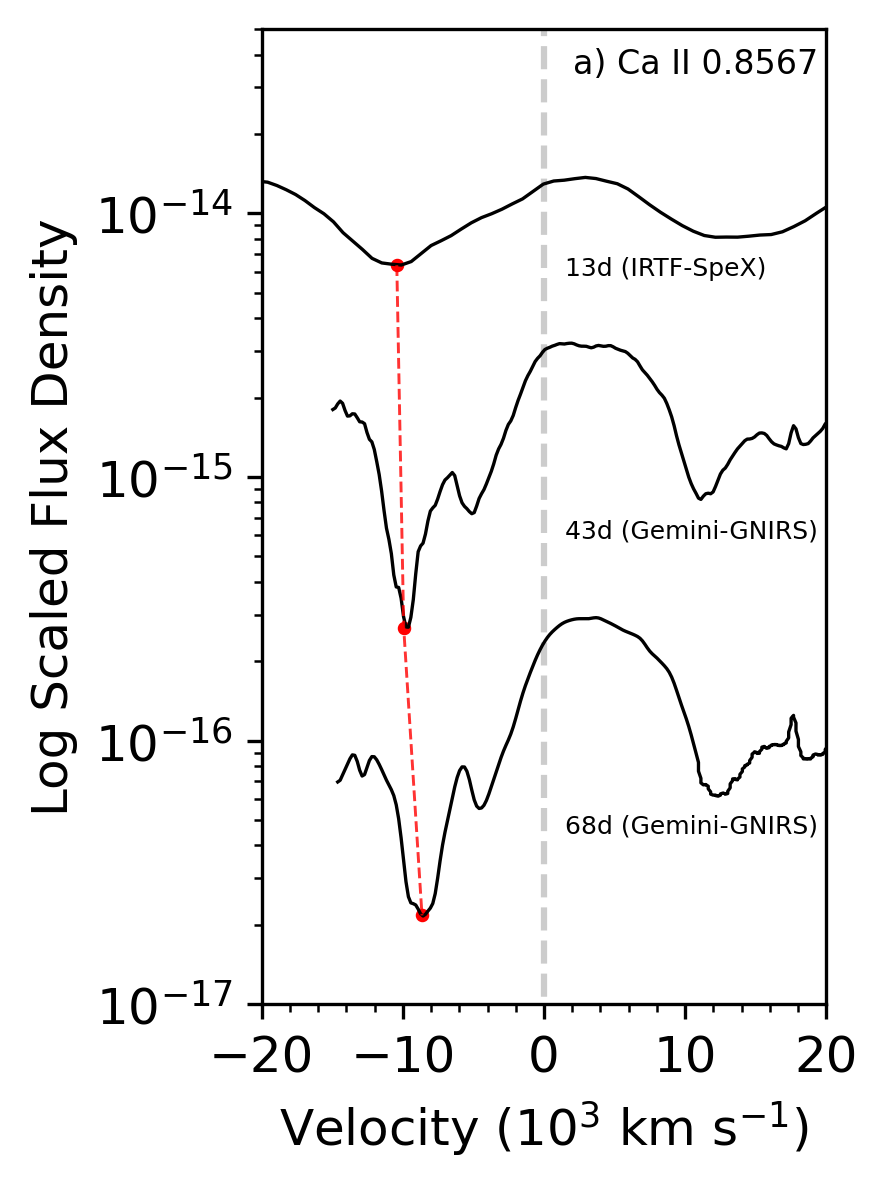} &
\includegraphics[scale=0.55,keepaspectratio]{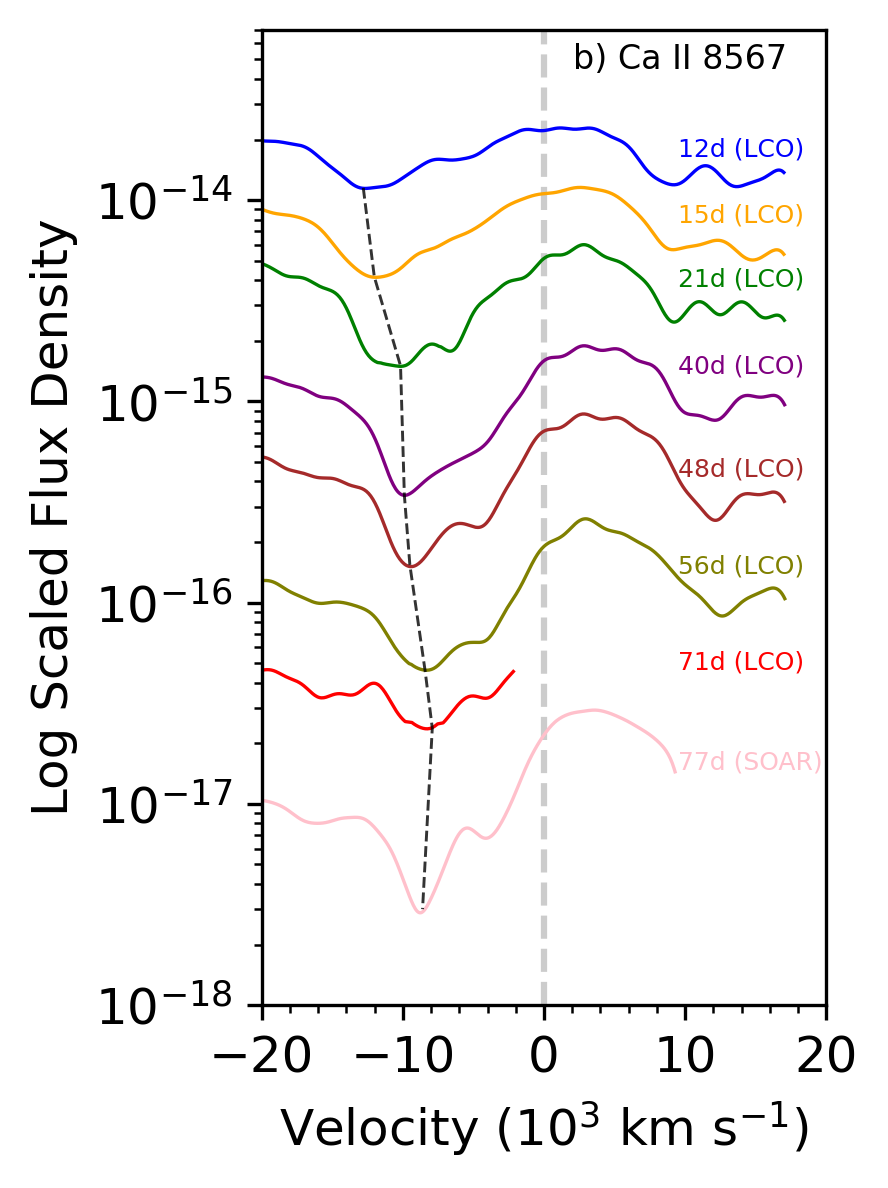} & 
\includegraphics[scale=0.55,keepaspectratio]{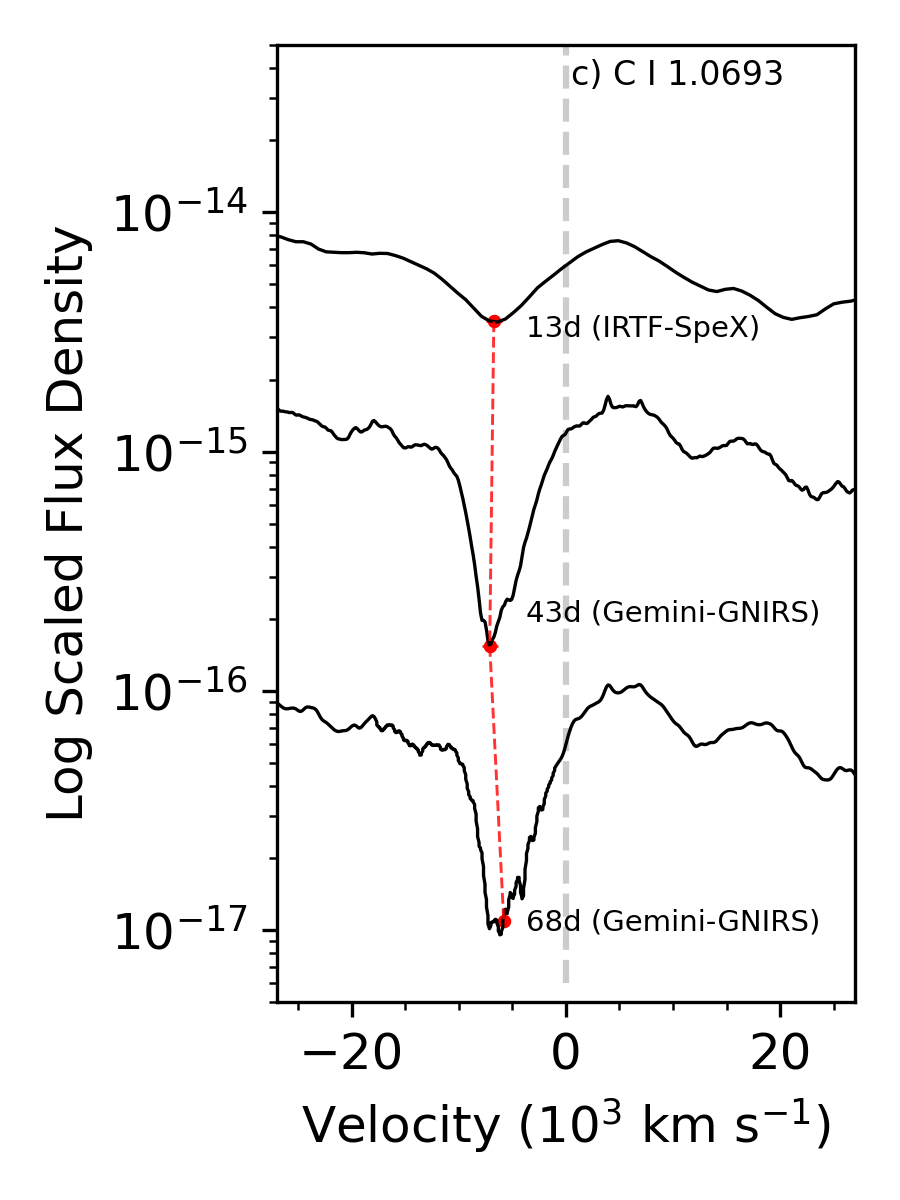} &
\includegraphics[scale=0.55,keepaspectratio]{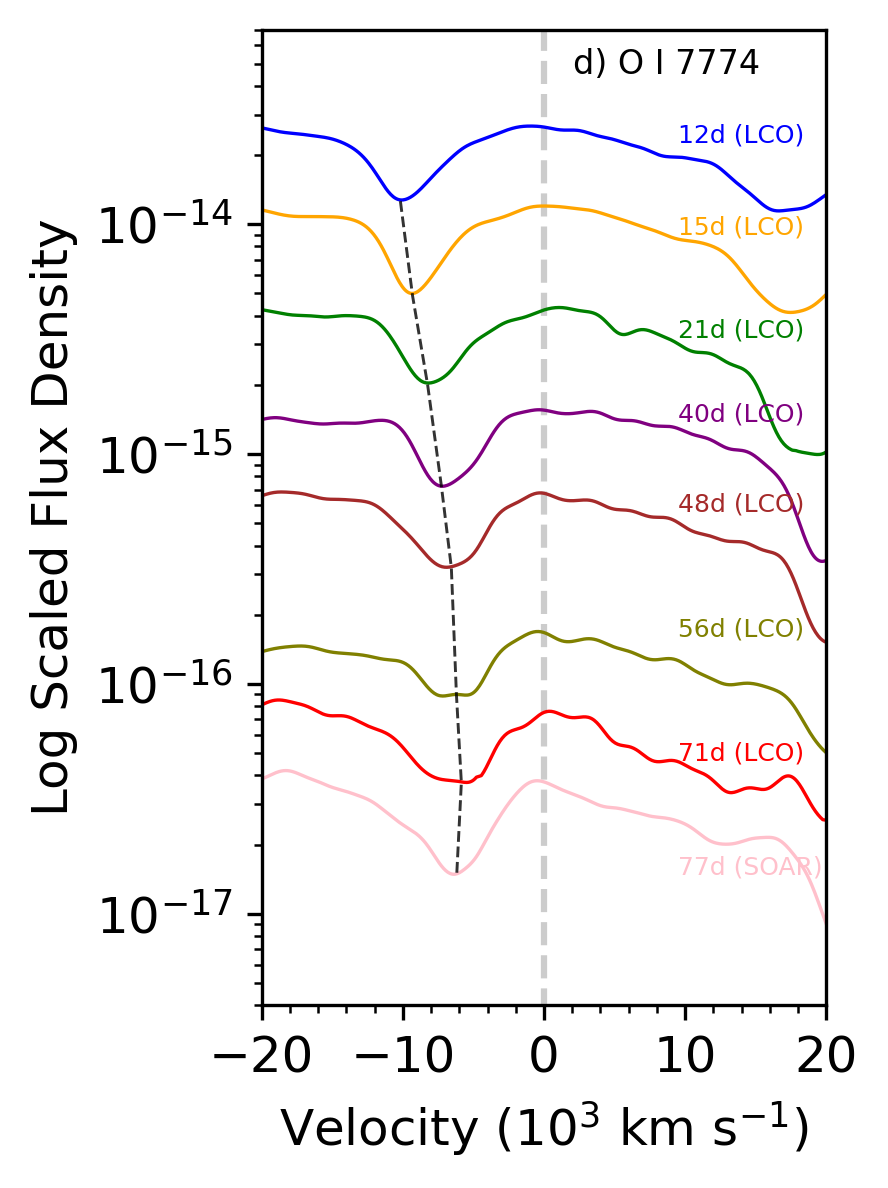} \\
\\
\end{tabular}

\caption{Evolution of the velocity profiles of the Ca~II triplet observed with NIR spectrographs (panel (a)) and the optical spectrographs at LCO and SOAR (panel (b)), the C~I 1.0693\,${\mu}$m line (panel (c)), and the O~I line at 7774\,\AA\ (panel (d)). The temporal shift of the absorption minima in panels (b) and (d) are marked with a black dashed line and their associated ion velocities are given in Table \ref{tab:vel}. The velocities of absorption minima for the NIR Ca~II triplet (panel (a)) are marked as red circles at velocities $-10,400$, $-9900$, and $-8600$\,km\,s$^{-1}$ at days 13, 43, and 68, respectively. Similarly, for C~I 1.0693\,${\mu}$m (panel (c)), the velocities associated with absorption minima are marked as red circles at $-6700$, $-7100$, and $-5800$\,km\,s$^{-1}$ at days 13, 43, and 68, respectively.}
\label{vel}
\end{figure*}

The velocity profiles of the strong absorption lines of Ca~II and O~I observed with the optical spectrographs at LCO and SOAR, along with the profiles of the S~I line and the same Ca~II lines observed with NIR spectrographs, are presented in Figure \ref{vel}. We estimate an ion velocity of $\sim 11,000$\,km\,s$^{-1}$ from the optical observation at day 12 for the Ca~II absorption minimum. As SYN++ models do not work well with NIR spectra owing to incomplete line lists, we employ simple Gaussian modeling to estimate velocities. Using three Gaussians to fit the NIR Ca~II triplet at day 13, we estimate an ion velocity of $10,400 \pm 200$\,km\,s$^{-1}$ from its absorption minima. This result is consistent with the independently derived optical velocities of $\sim 11,000 \pm 500$\,km\,s$^{-1}$ for Ca~II using SYN++. 

\begin{figure*} 
\centering
\begin{tabular}{cc}
\includegraphics[width=0.5\textwidth,keepaspectratio]{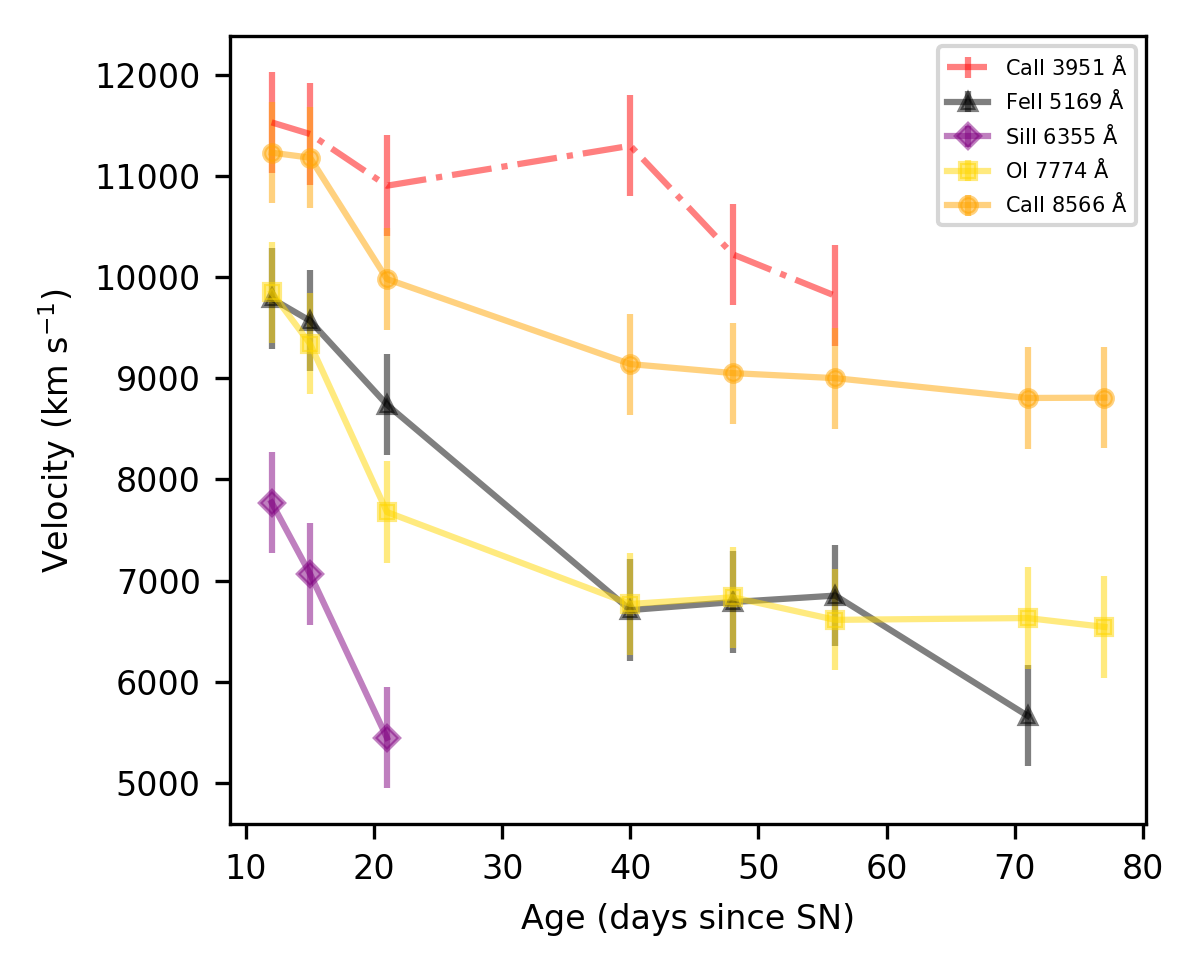} & 
\includegraphics[width=0.5\textwidth,keepaspectratio]{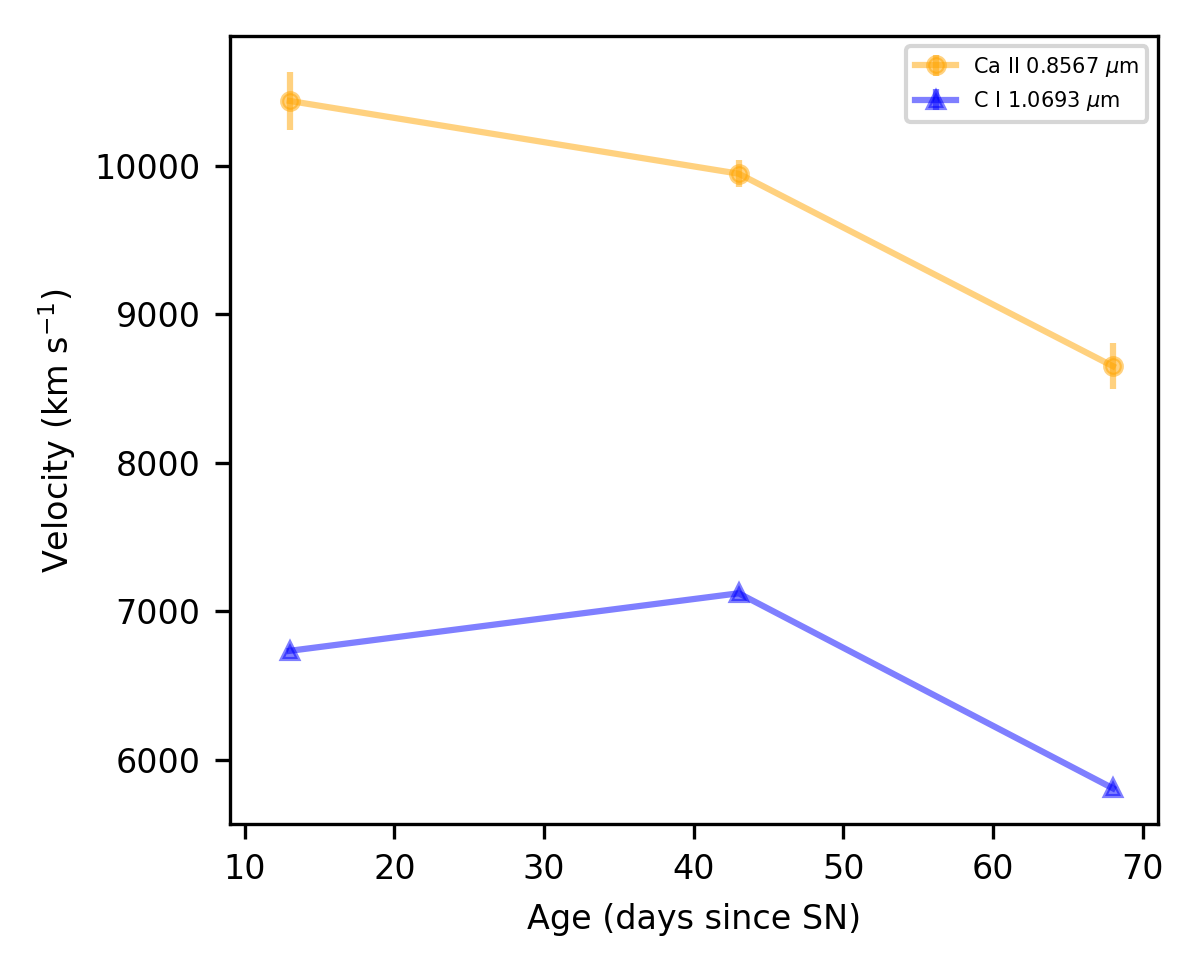}  \\
(a) & (b) \\
\end{tabular}
\caption{Velocities of some of the identified atomic lines are plotted as a function of time. In panel (a) the velocities were estimated using SYN++ modeling as discussed in Section \ref{sec:4.2} and presented in Table \ref{tab:vel}. In panel (b), the velocities were measured using multi-Gaussian modeling of the NIR spectra to identify Doppler shifts of the absorption minima from the rest wavelengths.}
\label{vel_vs_time}
\end{figure*}

The temporal evolution of ion velocities at optical and NIR wavelengths is shown in Figure \ref{vel_vs_time}. The lines of O~I, Ca~II, and Fe~II exhibit higher velocities than Si~II at the first three epochs (days 12, 21, and 40) when they are detected. This is probably due to the higher optical depths in their lines, which may cause only the outer higher-velocity layers of the homologously expanding ejecta to be observed. Differences between velocities of outer layers for Fe and lighter element such as Si~II \citep{hoflich91} could be indicative of asphericities in the ejecta distribution as is evident in some CCSNe \citep[e.g.,][]{mazzali01}. At NIR wavelengths, both Ca~II and C~I absorption minima decrease in velocity between days 13 and 68, consistent with a receding photosphere observed at the beginning of the nebular phase. 

\begin{figure} [h]
\includegraphics[width=0.5\textwidth]{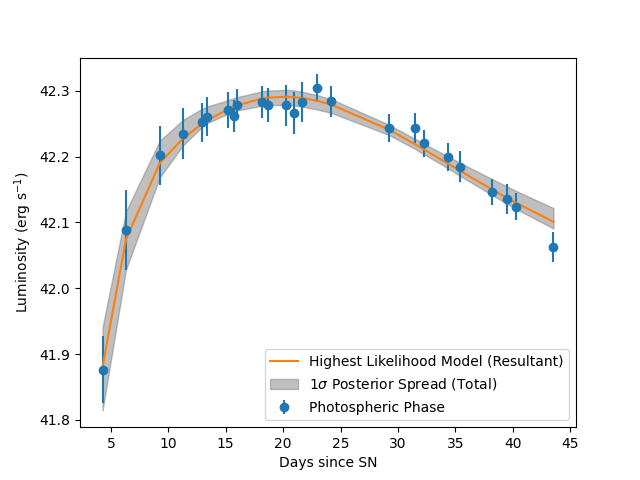}
\caption{Bolometric light-curve fit of photospheric phase ($t < 45$ days) with the Arnett model. The highest-likelihood model is presented as an orange curve and the corresponding 1$\sigma$ posterior spread is displayed in gray.}
\label{bolometric_fit_Arnett}
\end{figure}

\begin{figure} 
\centering
\begin{tabular}{c}
\includegraphics[width=7.8truecm]{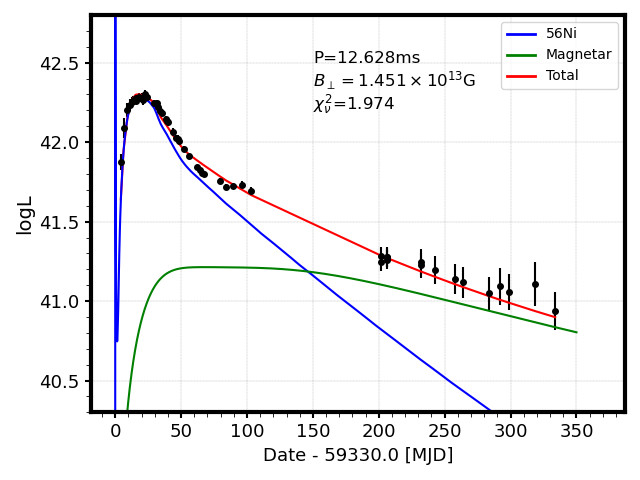} \\ 
(a)\\
\includegraphics[width=7.8truecm]{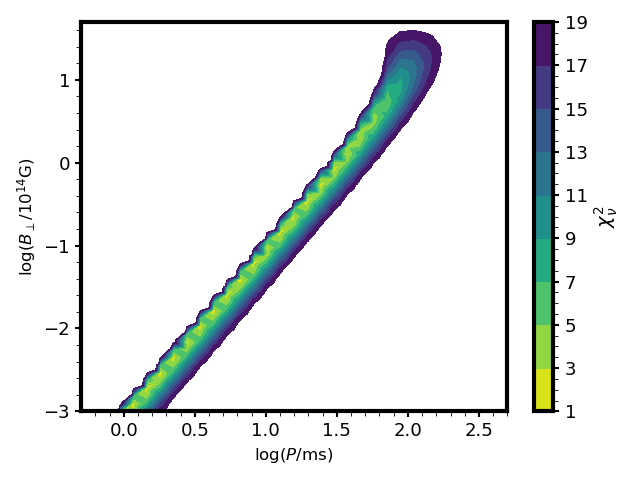} \\
(b)\\
\includegraphics[width=7.8truecm]{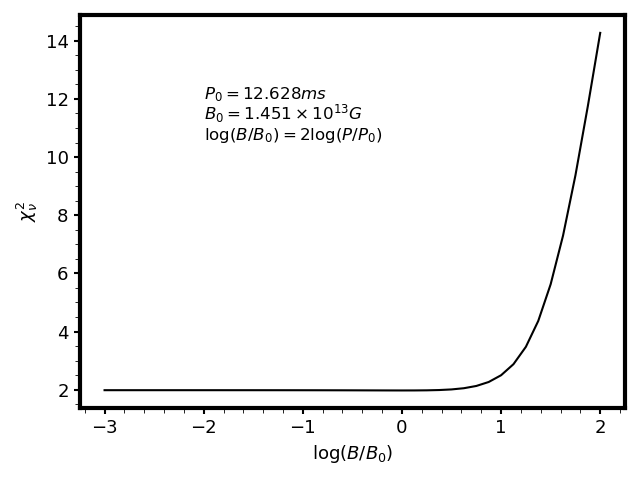} \\
(c)\\

\end{tabular}
\caption{(a) Best-fit model to the bolometric light curve combining radioactive decay ($M_\mathrm{Ni} = 0.10$\,M$_\odot$) along with an additional powering magnetar ($P_\mathrm{init} = 12.62$\,ms, $B_{\perp} = 1.45 \times 10^{13}$\,G). (b) Distribution of the two-component light-curve model fits. The 90\% and 99\% confidence contours are consistent with $\chi^{2}_{\nu} = 2.71$ (yellow) and 6.63 (green), respectively. A strong degeneracy between $B_{\perp}$ and $P_\mathrm{init}$ is shown in the ranges of $10^{11}$--$5 \times 10^{13}$\,G and 1--15\,ms, respectively. (c) The $\chi^{2}_{\nu}$ distribution as a function of magnetic fields relative to the best-fit value ($B_\mathrm{0}$). Magnetic fields are well constrained at $B_{\perp} < 5 \times 10^{13}$\,G, but unconstrained in the lower bound.}
\label{Full_Ni}
\end{figure}

\subsection{Bolometric Light Curve Modeling} \label{sec:4.3}

\subsubsection{Arnett Modeling: Photospheric Phase} \label{sec:4.3.1}
We fit the constructed bolometric light curve (see Section \ref{sec:3.2}) with the semi-analytic prescription of Arnett \citep[hereafter the ``Arnett Model'';][]{arnett82, valenti08a, chatzopoulos12, cano13} for the photospheric phase ($t < 45$ days). This model assumes a homologous expansion of uniform-density ejecta with no nickel mixing, a constant optical opacity (${\kappa}_\mathrm{opt}$), a small initial radius before explosion, and optically thick ejecta. We adopt the method of \citet[][see Eq. 1]{valenti08a}, where radioactive decay of both $^{56}$Ni and $^{56}$Co are assumed to be energy sources. As noted by \cite{lyman16}, we account for an incorrect numerical factor, 3/5 (see Eq. 2 of \cite{valenti08a}), instead of the correct fraction, 5/3, which propagates into the expression for the timescale of the light curve ($\tau_{m}$ in \cite{arnett82}). 

It has been noted for SN~Ic bolometric light-curve fitting that best fits are obtained when we schematically consider the ejecta contributions to luminosity from two regions: a high-density inner region and a low-density outer region \citep{maeda03, valenti08a}. We assume that the luminosities emitted by these two regions sum up to produce the overall luminosity when emission from the inner region is not absorbed by the outer counterpart. In the photospheric phase, the inner region is optically thick and its emerging luminosity is a small fraction of the total luminosity.  
\begin{table*}
\begin{center}
\caption{Best-fit and Grid Parameters of Neutron Star$^a$ and Ni Models }
\begin{tabular}{llccccccc}
\hline \hline
Model &$\chi^2_{\nu}$ & $E_\mathrm{k}$ & Ni mass & $B_{\perp}$ & $P_\mathrm{init}$ & $t_\mathrm{exp}$$^b$ \\
                  &               & ($10^{51}$\,erg) & (M$_\odot$) & (G)& (ms) & (MJD)\\ 
\hline

CO3.93\_E0.5\_Mni0.11 & 2.8 & 0.50 & 0.11 & $1.5 \times 10^{11}$ & 1.51  & 59330\\
CO3.93\_E0.6\_Mni0.10      &1.97 &0.61 &0.10 & 1.45 $\times$ 10$^{13}$ & 12.62 & 59330\\
                     
{CO3.93\_E0.5\_Mni0.05}                    &12.09 & 0.50 &0.05 &   $1.1 \times 10^{15}$    & 54.8& 59330\\

\hline
CO5.74\_E1.05\_Mni0.11 & 5.73 & 1.05 & 0.11 &   $1.0 \times 10^{11}$ & 1.31  & 59330\\
{CO5.74\_E1.44\_Mni0.10}                       &  3.89   & 1.44 & 0.10 & $5.3 \times 10^{13}$ & 24.16 & 59330 \\
CO5.74\_E1.44\_Mni0.05                      & 16.32  & 1.05 & 0.05 & $1.36 \times 10^{15}$ & 47.75 & 59330\\
\hline  
\hline
\label{central_engine_fitting}
\end{tabular}
\end{center}
\renewcommand{\baselinestretch}{0.8}
\tablecomments{$^a$ We consider either a normal millisecond pulsar or a magnetar.\\}
\end{table*}
The best-fit bolometric light curve is presented in Figure \ref{bolometric_fit_Arnett}. Assuming a constant opacity, ${\kappa}_\mathrm{opt}$ = 0.07\,g$^{-1}$\,cm$^{2}$ \citep{cano13, taddia2016iptf15dtg}, we obtain the following explosion parameters: nickel mass, $M_\mathrm{Ni} = 0.118 \pm 0.007$\,M$_{\odot}$; total ejecta mass, $M_{\rm ej} = 2.76 \pm 0.44$\,M$_{\odot}$; and explosion kinetic energy, $E_{\rm k} = (0.73 \pm 0.23) \times 10^{51}$\,erg. These are in reasonable agreement with those from our STELLA model fits (Table \ref{Tprogenitor} and Section \ref{sec:3.1}). 

We revisit the degeneracy between ejecta mass and explosion kinetic energy as described in Section \ref{sec:3.1} (Figure \ref{theory_models}) with additional constraints from spectroscopy. In the Arnett model, if one assumes a homogeneous density distribution of the ejecta, the measured photospheric velocity near peak luminosity ($v_{\rm phot}$) is related to $M_\mathrm{ej}$ and $E_\mathrm{k}$ through the relation \citep{arnett82}
\begin{equation}
    v_{\rm phot} \approx \frac{5}{3} \sqrt{\frac{2E_\mathrm{k}}{M_\mathrm{ej}}}\, .
\end{equation}
From {\tt SYN++} models as discussed in Section \ref{sec:4.2}, the measured photospheric velocity near peak luminosity (at day 21) is $7000 \pm 500$\,\kms. In model CO-3.93 (Table \ref{Tprogenitor}), for the best-fit values of  $M_\mathrm{ej} = 2.49$\,M$_{\odot}$ and $E_\mathrm{k} = 0.5 \times 10^{51}$\,erg, from the above equation we derive $v_{\rm phot} = 7500$\,\kms. On the other hand, for model CO-5.74 (Table \ref{Tprogenitor}), with corresponding best-fit parameters $M_\mathrm{ej} = 4.08$\,M$_{\odot}$ and $E_\mathrm{k} = 1.05 \times 10^{51}$\,erg, the derived velocity is $v_{\rm phot} = 8500$\,\kms. Thus, the combination of best-fit ejecta mass and explosion kinetic energy based on model CO-3.93 is in better agreement with the observed photospheric velocity near peak luminosity ($7000 \pm 500$\,\kms) than when using CO-5.74. We note that a better fit using CO-3.93 compared to CO-5.74 is also obtained when simultaneously modeling the early- and late-time bolometric light curve in Section \ref{sec:4.1}.

\subsubsection{Additional Power Source: Central Engine} \label{sec:4.3.2}
The $^{56}$Ni radioactive decay powering the peak luminosity, along with its decay product $^{56}$Co at late times, cannot reproduce the late-time observed light curves (Figure \ref{theory_models}). The Arnett and STELLA models give consistent results in the photospheric phase (Table \ref{Tprogenitor} and Section \ref{sec:4.3.1}). Thus, we hereafter use a combination of STELLA models and an additional power source to fit the entire bolometric light curve including early- and late-time data. We consider heating of the ejecta by magnetic dipole radiation from a central neutron star (e.g., magnetar or millisecond pulsar) as an additional power source to fit the late-time excess. In our modeling, we hereafter refer to the magnetic neutron star as the central engine, which may be either a millisecond pulsar with relatively low magnetic fields ($B < 10^{14}$\,G) or a magnetar with strong magnetic fields ($B \approx 10^{15}$\,G). This is because magnetar parameters at birth are not well constrained, thus making a wide parameter space of initial spin periods and magnetic fields possible depending on the properties of the progenitor system.

We considered a combination of a magnetar and nickel-decay ($M_\mathrm{Ni} = 0.11$, 0.10, and 0.05\,M$_{\odot}$). We employed Equations (2)--(7) of \cite{nicholl17} to calculate the bolometric luminosity added by magnetic dipole radiation heating from the neutron star. We constructed a two-dimensional parameter grid consisting of the initial spin period ($P_\mathrm{init}$) and the component of magnetic field perpendicular to the spin axis ($B_{\perp}$), and fitted it to the bolometric light curve from Section \ref{sec:3.2}. For the case with no nickel-decay contribution, we considered different onset dates of the central engine before the explosion date as a free parameter, $t_\mathrm{exp}$. We fixed the optical opacity, ${\kappa}_{opt} = 0.1$\,g$^{-1}$\,cm$^{2}$, the opacity to high-energy photons, ${\kappa}_{\gamma} = 0.1$\,g$^{-1}$\,cm$^{2}$, and the mass of the central neutron star, $M_\mathrm{NS} = 1.4$\,M$_{\odot}$.

Combining radioactive decay with an additional power source in our models, together with grid parameters assuming two progenitor models (CO-3.93 and CO-5.74; Section \ref{sec:3.1}), we obtain the results shown in Table \ref{central_engine_fitting}. We varied $B_{\perp}$ and $P_\mathrm{init}$ over the parameter spaces $10^{11}$--$5 \times 10^{15}$\,G and 0.5--500\,ms, respectively. Fixing $M_\mathrm{Ni}$ to the best-fit value ($M_\mathrm{Ni} = 0.11$\,M$_{\odot}$) found by our STELLA modeling (see Section \ref{sec:3.1} and Table \ref{Tprogenitor}), we find that the late-time excess of the bolometric light curve can be fitted by a pulsar with $B_{\perp} = 1.5 \times 10^{11}$\,G and an initial period $P_\mathrm{init} = 1.5$\,ms. However, owing to degeneracies, large ranges of $B_{\perp}$ and $P_\mathrm{init}$ ($B_{\perp} = 10^{11}$--$5 \times 10^{13}$\,G; $P_\mathrm{init} = 1$--15\,ms) are allowed. We note that the gamma-ray trapping efficiency, which is parameterized with ${\kappa}_{\gamma}$ in our magnetar models, is uncertain at late times.

In these composite models the CO-3.93 progenitor models systematically provide better fits than the CO-5.74 model (see Table \ref{central_engine_fitting}). A preference for CO-3.93 is also the case when constraining $M_\mathrm{ej}$ and $E_\mathrm{k}$ based on photospheric velocity at peak luminosity (see Section \ref{sec:3.2}). For $M_\mathrm{Ni} = 0.10$\,M$_{\odot}$, the statistical improvement in the best-fit model is significant, particularly to fit the late-time excess ($\chi^2_{\nu} \approx 1.9$; Table \ref{central_engine_fitting}, Figure \ref{Full_Ni}). Considering only CO-3.93, the best-fit parameters are $B_{\perp} = 1.45 \times 10^{13}$\,G and $P_\mathrm{init} = 12.62$\,ms. Although this model fit is statistically good, it still allows for large ranges of parameter values (see Figure \ref{Full_Ni}b, Figure \ref{Full_Ni}c). However, our modeling clearly favors relatively low magnetic fields ($B_{\perp} < 5 \times 10^{13}$\,G) to fit both the radioactive-powered peak and the late-time tail simultaneously.   

As discussed above, our best-fit model is not unique and should only be considered as indicative. To get a better-constrained solution, a full parameter search is necessary, with an extensive model grid varying several different parameters simultaneously, including $E_{\rm k}$, $M_\mathrm{Ni}$, $B_{\perp}$, $P_\mathrm{init}$, ${\kappa}_\mathrm{opt}$, ${\kappa}_{\gamma}$, etc., which is beyond the scope of this paper. Our model fits with $M_\mathrm{Ni} = 0.05$\,M$_{\odot}$ are significantly worse than fits for higher nickel masses (Table \ref{central_engine_fitting}), and are not able to reproduce the observed late-time fluxes. 

Despite the uncertainties in model fitting, our results suggest that at early times ($t < 50$\,days since the SN) the powering mechanism is dominated by radioactive decay, but at late times ($t > 200$ days) the contribution from the central engine becomes significant. While the detailed properties of the powering sources of SN 2021krf are uncertain, we can confidently conclude that the $^{56}$Ni and $^{56}$Co radioactive decay alone (Figure \ref{theory_models}) cannot adequately describe the observed light curves at both early and late times simultaneously. The central engine that is needed to account for the excess luminosity at late times must have a period on the order of milliseconds at birth and relatively low magnetic fields ($< 5 \times 10^{13}$\,G).

An alternate explanation for the observed late-time excess could be SN-CSM interaction. We discuss the possibility of such interaction in Section \ref{sec:4.4}. While this scenario might explain both the observed light curve and rising NIR continuum, the lack of direct evidence of such interaction discourages us from favoring this explanation.

\subsection{Origin of Dust Emission} \label{sec:4.4}

NIR emission from dust was directly detected on day 68 based on the rising continuum longward of 2.0\,$\mu$m. In general, the possible origins and locations of the dust in CCSNe are (a) formation of dust in the expanding SN ejecta; (b) formation of dust in the dense CSM surrounding the SN; and (c) radiative heating by the SN flash of pre-existing dust in the surrounding CSM/ISM, the so-called ``IR echo.'' We discuss each of these possibilities below.\\

\textbf{\underline {Dust formation in the SN ejecta:}} Since no IR spectra were obtained after day 68, we searched for the presence of dust in optical spectra to examine the origin of the dust emission in SN 2021krf. The evolution of emission-line profiles in the nebular phase of the SN ejecta can reveal the presence of newly formed dust. Newly condensed dust may obscure the receding ejecta, suppressing the redshifted component of emission lines and resulting in asymmetric emission-line profiles. Indeed, dominant blue wings on hydrogen and helium emission lines are often observed in CCSNe \citep{elmhamd04, smith08}. 
\begin{figure}
\includegraphics[width=0.5\textwidth]{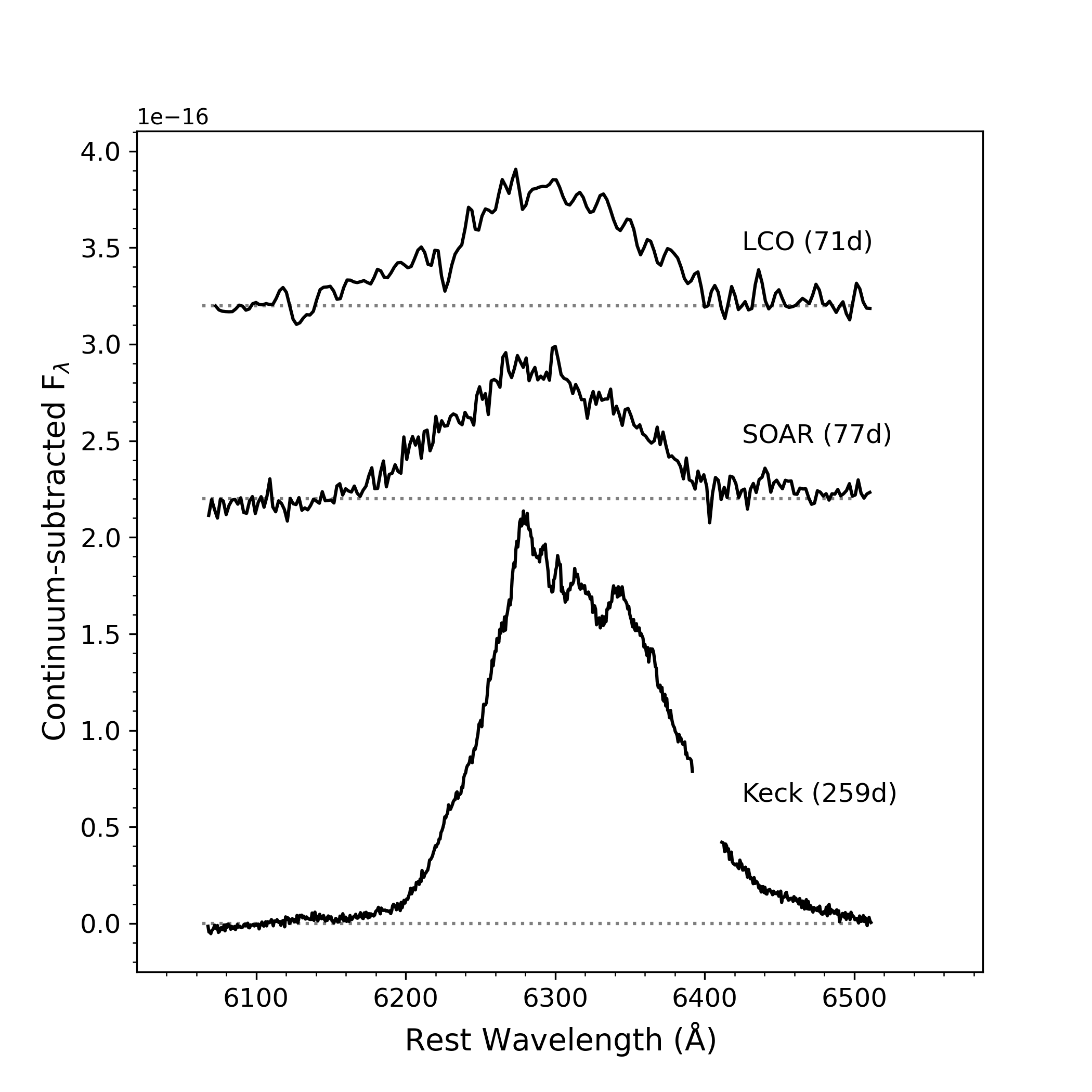} \\
\caption{Scaled and continuum-subtracted [O~I] $\lambda\lambda$6300, 6364 line profiles on days 71, 77, and 259. The gap in the spectrum on day 259 is due to a lack of data at these wavelengths. Grey dotted lines indicate the continuum level for each spectrum.}
\label{OIdoublet}
\end{figure}

\begin{figure}
    \centering
    \includegraphics[width=0.5\textwidth]{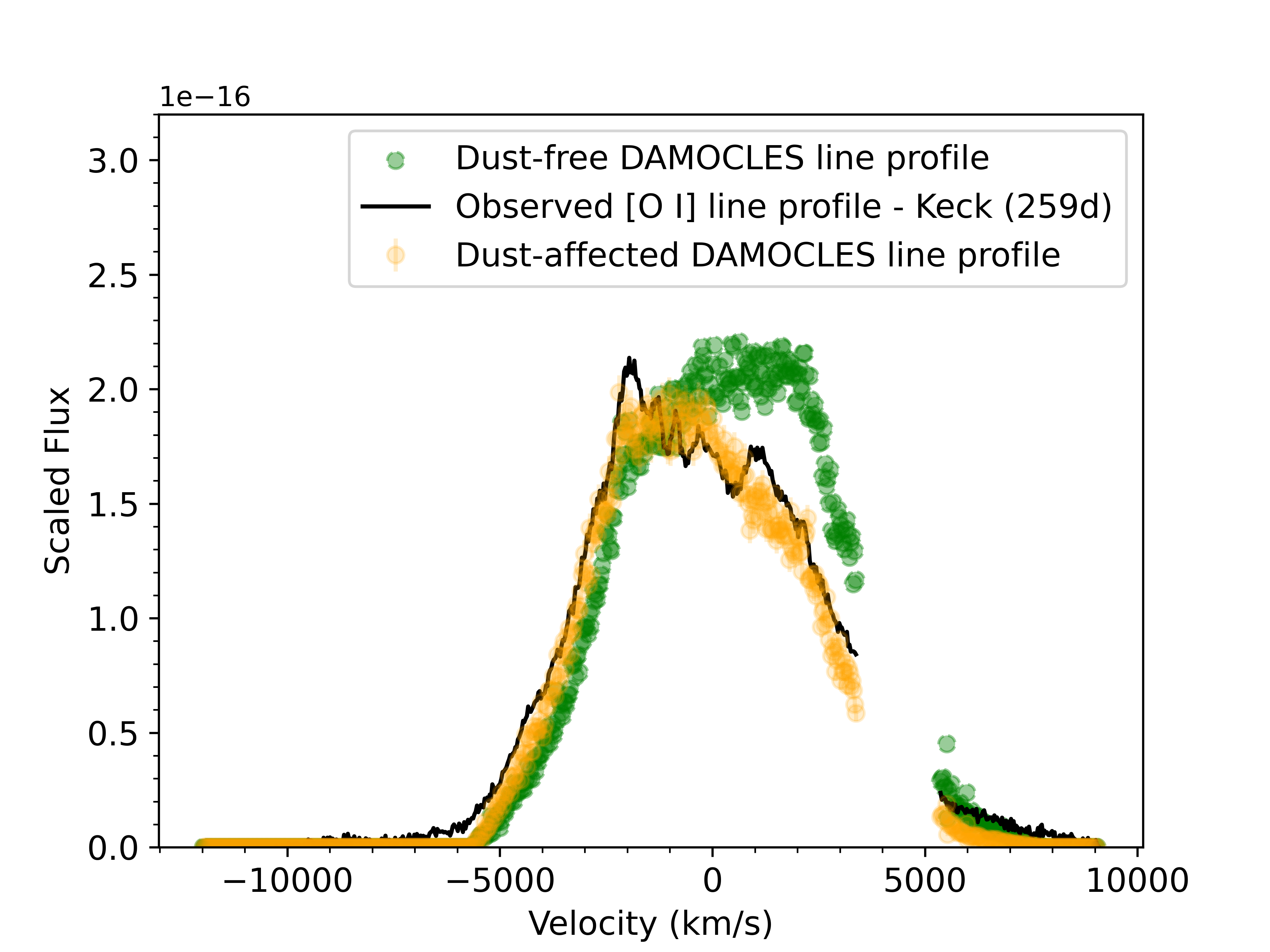}
    \caption{DAMOCLES models of the [O~I] $\lambda\lambda$6300, 6364 line profile in SN 2021krf with (orange) and without dust (green) contributions. The observed nebular line profile at day 259 is shown in black. A smooth amorphous carbon-grain dust model is assumed. Modelled line profile with a dust mass of (2.1 $\pm$ 1.5) $\times$ 10$^{-5}$ M$_{\odot}$ and a grain size of 0.4 $\pm$ 0.1\,$\mu$m best describes the observed line profile.}
    \label{asymmetry_blue_red}
\end{figure}

In the stripped-envelope SN 2021krf, strong aysmmetry is present in the blended [O~I] $\lambda\lambda$6300, 6364 lines from the nebular-phase optical spectrum at day 259 (Figure \ref{Foptispec21krf}). The detailed emission profile of this blended line complex is shown in Figures \ref{OIdoublet} and \ref{asymmetry_blue_red}. We note hints of asymmetries in the [O~I] profile at significantly earlier epochs (e.g., days 71, 77 --- Figures \ref{Foptispec21krf}, \ref{OIdoublet}). However, the SNR of the [O~I] emission profile at these epochs is significantly lower ($\sim$ 1.3--2) than the SNR on day 259 ($\sim$ 10). Also, between days 71 and 77 the observed optical SN spectra are still not completely nebular, as several strong absorption lines are present (Figure \ref{Foptispec21krf}). Hence, we focus on the asymmetry at day 259 when the observed optical spectrum is truly nebular.

Studies of the [O~I] $\lambda\lambda$6300, 6364 doublet emission lines in stripped-envelope SNe have shown a predominance of blueshifted features \citep{taubenberger09, milisavljevic10}. Internal scattering or dust obscuration of the emission from the far-side ejecta were suggested to be the most likely causes of the asymmetry by \cite{milisavljevic10}. However, based on a sample of 39 SNe~Ib/c, \cite{taubenberger09} favored the opaque-ejecta scenario to explain the observed predominantly blueshifted peaks. Thus, the relation of blueshift of the [O~I] doublet to dust formation is debatable. On the other hand, in a large sample of CCSNe as discussed by \cite{niculescu-duvaz22}, ejecta emission show an asymmetric red scattering wing and/or a blueshifted peak, both of which can be caused by dust absorption and scattering. Double-peaked [O~I] emission in the late-time optical spectra of CCSNe have been suggested to be associated with O-rich ejecta \citep[e.g.,][]{maeda02, mazzali05, modjaz08, taubenberger09, milisavljevic10}. Hence, asymmetric [O~I] line profiles in the late-time spectra of stripped-envelope SNe can be indicators of actual physical differences in the emission from the blueshifted and redshifted expanding SN ejecta.

Motivated by the statistical significance of blue--red flux asymmetries in CCSNe, we have compared the observed [O~I] $\lambda\lambda$6300, 6364 emission-line profile of SN 2021krf with synthetic line profiles constructed using the Monte Carlo radiative transfer code DAMOCLES \citep{bevan16}. DAMOCLES  models the effects of dust on the line profiles of CCSNe in order to determine newly formed dust masses. We constructed models both with and without the presence of dust in the SN ejecta. We adopted a similar manual fitting procedure between synthetic and observed ejecta line profiles to that adopted by \cite{niculescu-duvaz22}. We iterated over grain radius and dust-mass parameters to find the best-fit DAMOCLES model.

\begin{figure*}
\includegraphics[width=\textwidth]{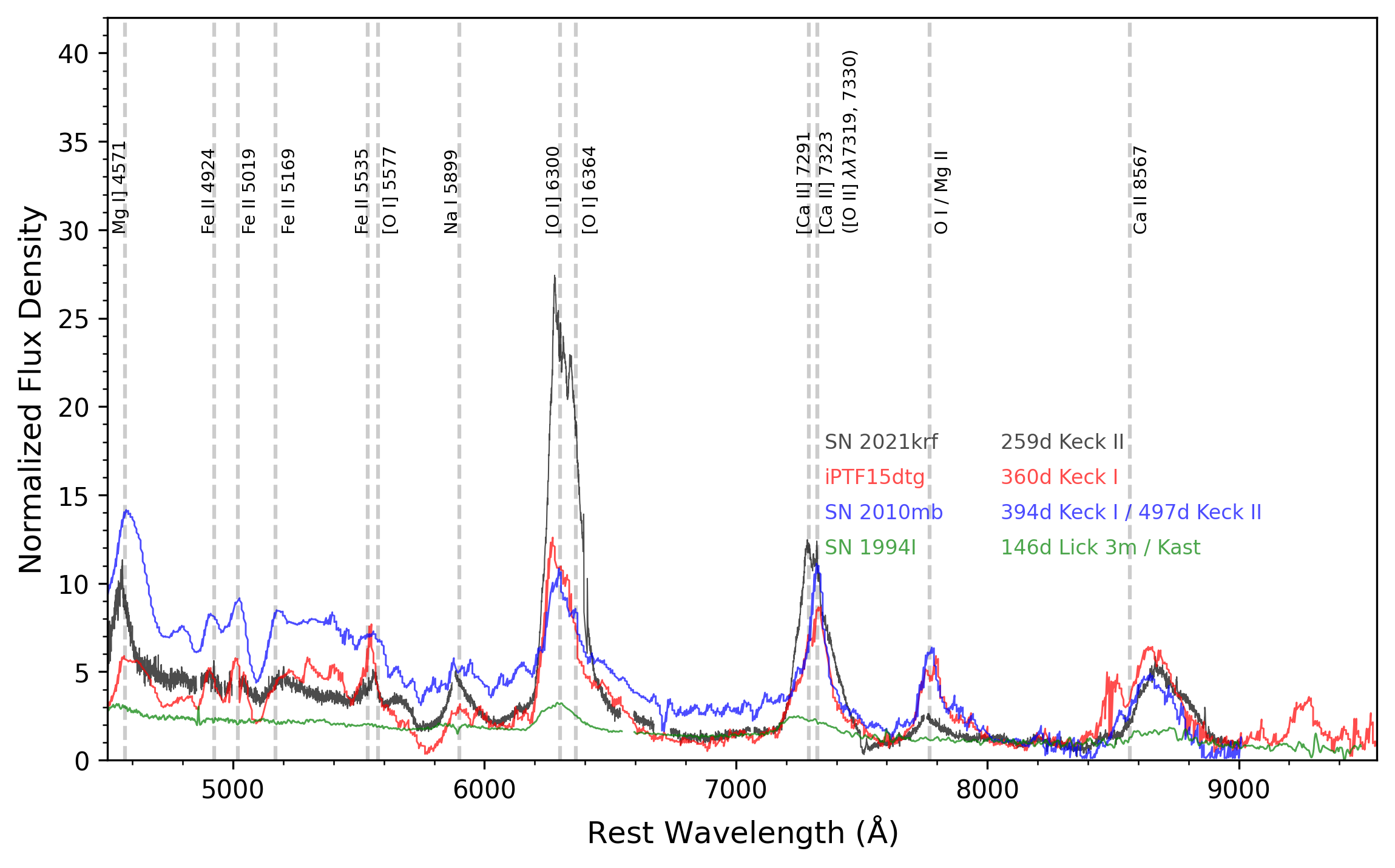}
\caption{The late-time spectrum of SN 2021krf on day 259 (black). For comparison, similarly late-time spectra from three other kinds of SNe~Ic (SN 1994I, SN 2010mb, and iPTF15dtg) are overlaid. All spectra have been deredshifted and corrected for extinction. Narrow lines from respective host galaxies are subtracted. Spectra of all SNe have been normalized (divided by their median fluxes between 8000 and 8400\,\AA) and the main spectral features have been marked. Potential blending of the [Ca~II] doublet with [O~II] $\lambda \lambda$7319, 7330 is noted in parentheses.}
\label{Late-time spectrum}
\end{figure*}

The best-fitting model ($\chi^{2}_{\nu} \approx 0.9$) as shown in Figure \ref{asymmetry_blue_red} has an associated dust mass of $(2.1 \pm 0.4) \times 10^{-5}$\,M$_{\odot}$ and a dust grain size ($a$) of $0.4 \pm 0.1$\,$\mu$m. The best-fitting dusty model is presented with the observed line profile at day 259 in Figure \ref{asymmetry_blue_red}. The dust-free DAMOCLES model is clearly an unacceptable fit to the observed line profile. The rising NIR continuum observed at day 68 (Figure \ref{FnearIRspec21krf}) suggests that dust formation could have started as early as day 68, continuing into the late-time epoch of day 259. The independently estimated dust-mass range ($\sim 0.5$--3.0 $\times 10^{-5}$\,M$_{\odot}$) from day 68 (Table \ref{dustemission_table}) is comparable to the amount of dust estimated from DAMOCLES modeling of the spectrum at day 259. The estimated large dust-grain size (on the order of 0.1\,$\mu$m) in SN 2021krf is also consistent with the recent theoretical size distribution prediction of \cite{marassi19} for amorphous carbon grains formed in CCSNe. Similar grain sizes have been previously discussed for dust formed in other CCSNe \citep[e.g.,][]{sarangi15, sluder18, priestley19}.  

Another clearly detected emission feature in the late-time spectrum of SN 2021krf is the [Ca~II] $\lambda\lambda$7291, 7323 doublet (Figure \ref{Late-time spectrum}). However, the doublet is significantly less asymmetric and weaker in flux compared to [O~I] $\lambda\lambda$6300, 6364 (Figure \ref{Late-time spectrum}). At late times, a flux ratio of [O~I] to [Ca~II] $> 1$ is expected in SNe~Ic \citep[e.g.,][]{fang22}. We investigated the [Ca~II] profile in SN 2021krf and found that its skewed shape is likely due to blending of [O~II]  $\lambda \lambda$7319, 7330 with [Ca~II], as previously noted for other SNe~Ic \citep[e.g.,][]{taubenberger06}. The presence of this contamination makes it difficult to statistically distinguish between the dusty and no-dust DAMOCLES line profile model fits of [Ca~II]. Also, from model nebular spectra of SNe~Ic (at day 200), \cite{Dessart21} show that for a range of progenitor models, the [Ca~II] doublet (in contrast to the [O~I] doublet) is not a very useful diagnostic of the progenitor (and thus the ejecta) or explosion properties, partly owing to blending effects. Thus, we do not discuss DAMOCLES modeling of the [Ca~II] line profile at day 259.

Our results suggest that dust formation in the SN ejecta is a plausible candidate to explain the observed asymmetry in the [O~I] emission line of SN 2021krf. To further constrain the nature of this dust emission (including its origin), it is crucial to have several additional spectra at significantly later epochs to study the temporal evolution of the [O~I] line profile and quantify its asymmetries.

In Figure \ref{Late-time spectrum}, we compare the late-time optical spectrum of SN 2021krf with the nebular-phase spectra of various SNe~Ic: (1) iPTF15dtg, in which a late-time excess (see Figure \ref{FLCcomp1}) was observed owing to magnetar powering \citep{taddia19}; (2) SN 2010mb, a peculiar SN~Ic with late-time excess showing evidence of SN-CSM interaction \citep{benami14}; and (3) SN 1994I, a spectroscopically typical SN~Ic \citep{filippenko95}. The late-time spectrum of iPTF15dtg is similar to that of SN 2021krf, also showing asymmetries in the [O~I] $\lambda\lambda$6300, 6364 emission-line profile. We find that a dust model synthesized with DAMOCLES, similar to SN 2021krf, can explain the observed asymmetry at day 360 in iPTF15dtg. However, we note that, unlike SN 2021krf, no NIR rising continuum as independent evidence of dust formation has been observed in iPTF15dtg. However, this does not necessarily rule out dust formation in this SN. In contrast to SN 2021krf and iPTF15dtg, SN 2010mb shows a significantly weaker asymmetry in the [O~I] emission doublet. The [O~I] feature in SN 1994I is weak (Figure \ref{Late-time spectrum}), and it is difficult to characterize its line profile. Thus, we did not model the [O~I] line profiles in SN 2010mb and SN 1994I. Based on spectral comparisons of these SNe~Ic, below we investigate possible scenarios for the CSM origin of the rising dust continuum in SN 2021krf.\\


\textbf{\underline{Dust from CSM}}: The change in the slope of the $K$-band continuum could also be due to radiative heating of pre-existing circumstellar dust produced by the SN progenitor or by newly formed dust in the swept-up dense CSM. In Type IIn SNe (e.g., SN 2005ip and SN 2006jd), \cite{fox09,fox10,fox11} found compelling evidence of continuum emission from warm dust in the dense CSM within $\sim 100$ days after the explosions. The mass of dust formed in dense CSM knots of the ejecta of these SNe were found to be small as expected, since the mass-loss rate is rather low. The long duration of the NIR excess was interpreted to be due to heating associated with radiative shocks forming at interfaces of the SN-CSM interactions \citep{fox09}. 

In the case of SN 2005ip, \cite{fox10} derived dust temperatures of 900--1100\,K and a dust mass of $\sim 5 \times 10^{-4}$\,M$_{\odot}$. They suggested that the emission originated either in newly formed dust in the ejecta or in a  dense cool circumstellar shell in which pre-existing CSM was being continuously heated by interactions with the ejecta. For the case of SN 2006jd, \cite{stritzinger12} estimated a warm-dust mass of (0.7--9.8) $\times 10^{-4}$\,M$_{\odot}$.  In addition to the possibility of early-time warm dust emission, both SNe 2005ip and 2006jd were found to exhibit thermal emission associated with a colder dust component ($T \approx 400$--500\,K). This dust emission has been attributed to SN-CSM interactions \citep{fox10,fox11} at later times. Note that Type IIn SNe have clear evidence for the existence of a circumstellar shell showing up as narrow lines of H and He during the early stages of these SNe.

In contrast to Type IIn SN spectra, SN 2021krf does not exhibit narrow H or He lines, which are usually considered as evidence of CSM. A late-time Keck spectrum of SN 2021krf exhibits strong [O~I] and [Ca~II] lines (Figures \ref{Foptispec21krf}, \ref{Late-time spectrum}), similar to those of iPTF15dtg \citep{taddia19}. Thus, it is unlikely that at late times there is any ejecta interaction with a H- and He-rich CSM in SN 2021krf. We conclude that dust emission originating from a typical CSM interaction like in SNe~IIn is unlikely in SN 2021krf.

At wavelengths between $\sim 4500$ and 6000\,\AA, the continuum of SN 2010mb exhibits a strong blue excess (see Figure \ref{Late-time spectrum}). The evolving blue emission observed in the optical spectra also showed a corresponding slow light-curve decline at late times \citep{benami14}. As its spectrum showed no H lines, \citet{benami14} utilized a H-poor SN-CSM interaction model to fit the unusually strong blue quasi-continuum, thus interpreting it as a product of the interactions of the ejecta with the surrounding CSM.  The corresponding slow light-curve decline at late times in SN 2010mb was also attributed to the interaction with a H-poor CSM. 

At late times, SN 2021krf has a significantly weaker (blue excess) optical emission (in the range $\sim 4500$--6000\,\AA) compared with SN 2010mb (Figure \ref{Late-time spectrum}). Optical emission at these wavelengths in SN 2021krf is similar to that of iPTF15dtg, where no obvious signatures of CSM interaction were found \citep{taddia19}. Also, model spectra of SNe~Ic are expected to have significant contributions from numerous very narrow P~Cygni profiles, mainly of Fe~II ejecta lines, especially at $\lambda < 5000$\,\AA\ \citep{dessart12a}. Through SYN++ modeling (see Section \ref{sec:4.1}), we have shown that the optical spectra of SN 2021krf at these wavelengths have strong Fe~II and Sc~II P~Cygni lines (Figure \ref{fig:57}). Thus, at least part of the blue flux observed in SN 2021krf, is due to these ejecta lines. Additionally, the evolution of a noticeable blue quasicontinuum is not observed in the optical spectra of SN 2021krf near the detected dust emission NIR epoch at day 68, which one would expect if CSM interaction were the origin of dust emission. These observations suggest that the optical spectroscopic evolution of SN 2021krf is more similar to that of other spectroscopically normal stripped-envelope SNe than SN 2010mb, and that it is unlikely that there is any significant ejecta-CSM interaction or circumstellar emission.

While there is no clear evidence from late-time spectra for H-rich or H-poor CSM interactions, all possible cases of CSM interaction and CSM dust emission cannot be ruled out. As discussed in Section \ref{sec:4.3.2}, interaction of the ejecta with the CSM in SN 201krf would also explain the observed late-time luminosity excess. However, owing to a lack of clear observational evidence, we do not consider CSM interaction as the most likely source of dust emission in SN 2021krf. A more detailed discussion of the possible existence of other CSM signatures in SN 2021krf is beyond the scope of our current work.\\

\textbf{\underline{Thermal IR Echo}}: A third possible explanation for the NIR excess observed in SN2021krf is an IR echo from pre-existing CSM dust. This pre-existing dust is likely formed in the progenitor’s wind.  In this scenario, the SN explosion radiatively heats pre-existing dust lying beyond a dust-free cavity, producing an IR echo \citep[see][]{bode80, bode80snII, dwek83, Emmering88}.  NIR excesses around some SNe~IIn have been explained by formation of circumstellar shells, due to IR echoes around them \citep{dwek83, graham86}. Such circumstellar shells often cause a relatively high extinction toward SNe \citep{graham86}. 

SN 2021krf is a Type Ic SN and has relatively low extinction (see Section \ref{sec:3.3} and Figure \ref{extinction}). Moreover, until day 259, the spectra of SN 201krf show no clear evidence of CSM interaction which could make a dust-free zone for the IR echo. For the NIR excess emission observed at 68 days to be consistent with an IR echo, existence of CSM outside the dust-free cavity should be established. However, in optical spectra at day 71 and thereafter until day 259, no signs of SN-CSM interaction are found (Figure \ref{Foptispec21krf} and Figure \ref{Late-time spectrum}). Because the late-time spectrum (at 259 days) is likely CSM-free, the dust-free cavity formed by a light echo would have to be at least larger than 259 light days in radius (i.e., $\sim 10^{17}$\,cm).
Based on our current observations, this would be a crude lower limit for the cavity size, if an IR echo had produced it. For a massive (20--30\,M$_\odot$) W-R progenitor star, the radius of the circumstellar shell is expected to be on the order of $10^{17}$ to $10^{19}$\,cm \citep{garcia-Segura96}. Additionally, such a cavity would have to be dust-free, whereas the NIR excess in SN 2021krf was found significantly earlier, at day 68.  
Thus, contributions to the NIR excess by pre-existing dust in the CSM beyond the dust-free cavity through a thermal-IR echo is unlikely.\\

In summary, based on our observations, we suggest that among the three generally observed mechanisms for early-time (day 68) NIR excesses in SN 2021krf spectra, dust formation in the SN ejecta is the most likely. However, we cannot completely rule out the possibility of CSM interaction as the
cause of both the rising NIR continuum and the optical light-curve excess at late times.
Additional NIR and MIR spectra near and after day 68 would have better constrained the properties of the dust, its evolution, and its origin. IR spectra at later times than reported here are critical to unambiguously understand the dust emission in SNe~Ic. 

Although a CSM origin for dust cannot be completely ruled out, the spectra of SN 2021krf reported here, together with previously published observations of SN 2020oi, support early dust formation in SNe~Ic, probably in the ejecta. Dust features observed at near- and mid-infrared wavelengths, crucial for advancing our understanding of dust formation, evolution, and destruction in CCSNe, will be investigated at greater detail and to later (fainter) stages of evolution in the era of the {\it James Webb Space Telescope (JWST)} and the Extremely Large Telescopes (ELTs).

\section{Conclusion} \label{sec:5}

We have obtained NIR and optical observations of the Type Ic SN 2021krf in the galaxy 2MASX J12511712+0031138 (distantce 65\,Mpc) at the  Gemini North Telescope, the NASA Infrared Telescope Facility, the Las Cumbres Observatory, the Southern Astrophysical Research Telescope,  the Keck II Telescope, and the 3\,m Shane telescope at Lick Observatory.  From this work, we present the following conclusions.

\begin{enumerate}

\item SN 2021krf has relatively slower rising and declining times at early times ($t < 100$ days) than other SNe~Ic (e.g., SN 2007gr, SN 2020oi, and SN 1994I). Those rates are faster than in SN 2011bm and iPTF15dtg however, suggesting that SN 2021krf has a relatively low value of $E_\mathrm{k}/M_\mathrm{ej}$. The light curves of SN 2021krf at late times ($t > 200$\,days) decline slower than other typical SNe~Ic and $^{56}$Co radioactive decay, indicating the existence of an additional power source.

\item Calculations performed using the one-dimensional multigroup radiation hydrodynamics code STELLA imply that SN 2021krf has two degenerate solutions characterized by C-O star masses of 3.93 and 5.74\,M$_\odot$, but with the same best-fit nickel mass of 0.11\,M$_\odot$. The broad light curves indicate a low $E_\mathrm{k}/M_\mathrm{ej} \approx 0.2$.  The C-O star masses of 3.93 and 5.74\,M$_\odot$ correspond to $E_\mathrm{k}$ of 0.5 and $1.05 \times 10^{51}$\,erg and $M_\mathrm{ej}$ of 2.49 and 4.08\,M$_\odot$, respectively. The models fit the observed light curve well at $t < 40$ days, but do not at $t > 70$ days. Our best-fit models as well as Arnett modeling (Figure \ref{bolometric_fit_Arnett}) indicate that the photospheric phase ($t < 45$ days) of the light curves is well explained by radioactive decay alone. 


\item 
Optical spectroscopic monitoring shows dominant P~Cygni line profiles of Ca~II and clear [O~I] emission lines from day 71, progressively growing stronger until day 259, indicating the existence of significant amounts of O and Ca ejecta as expected from a stripped-envelope SN. Several Fe~II lines appear $\sim 21$ days after the SN. Emission in the Na~I doublet appears roughly at day 40 and increases in strength thereafter. NIR spectra show a strong Ca~II triplet, O~I, and C~I absorption lines and weak lines of Si~I and Mg~I. No clear evidence of He~I lines is observed.

\item Using the {\tt SYN++} code to model the optical spectra of SN~2021krf, we estimate a photospheric velocity of 10,000\,km\,s$^{-1}$ at day 12, a typical value of Type Ic SNe at early, pre-maximum phases. Significant temporal changes occurred between days 12 and 71, mainly from strong absorption due to O~I and Ca~II. The primary contributing lines to the spectra at day 71 are Ca~II, O~I, Fe~II, Sc~II, and Na~I. Doppler shift velocities for individual ions are in the range 8000--12,000\,\kms at day 12 depending on the line elements, and they decrease to 6000--10,000\,\kms by day 71.

\item A rising continuum longward of 2\,\mic\ was observed at day 68, as well as evidence for overtone CO emission. Fits to the continuum (2.0--2.3\,\mic) using carbon-dust and a grain size of 0.01, 0.1, and 1\,\mic\ yield dust masses of 2.4, 2.7, and $0.5 \times 10^{-5}$\,M$_{\odot}$ and corresponding dust temperatures of 900, 880, and 1170\,K, respectively. The estimated dust mass assuming MgSiO$_3$ dust is $2.8 \times 10^{-5}$\,M$_{\odot}$ with a dust temperature of 1020\,K.

\item To explain the rising continuum at day 68, we explored the possibilities that the dust is freshly formed in the ejecta, heated CSM dust, or an IR echo in the pre-existing CSM. Strong asymmetries in the [O~I] $\lambda\lambda$6300, 6364 line profile in the nebular spectrum at day 259 along with the NIR excess at day 68 support formation of dust in the SN ejecta. The lack of H and He lines, the low extinction toward SN 2021krf, and a lack of any strong quasi-continuum in the late-time optical spectrum suggests that interaction taking place between the ejecta and a pre-existing circumstellar shell which could cause either emission from heated CSM dust or an IR echo from the circumstellar shell is unlikely. The apparent lack of circumstellar shell in SN 2021krf, an SN~Ic, is in contrast to typical Type IIn SNe.

\item We modeled the heating of the ejecta by magnetic dipole radiation from a central neutron star (e.g., magnetar or millisecond pulsar) as an additional power source to explain the late-time excess in the bolometric luminosity. Neither radioactive decay nor a magnetar as a sole power source can explain all of the observed features of the light curves. We reproduce the bolometric light curve with a combination of radioactive decay ($M_\mathrm{Ni} = 0.10$\,M$_\odot$) and an additional powering source in the form of a central engine with an initial period on the order of milliseconds (12.62\,ms) and relatively low magnetic fields ($1.45 \times 10^{13}$\,G); the latter explains the late-time excess in the light curve between days 200 and 350.\\

\end{enumerate}

We thank the anonymous referee for helpful comments that improved this manuscript. We are grateful to Matt Nicholl for his insightful discussions on generating bolometric luminosity, and Griffin Hosseinzadeh for his help with the LCO photometry pipeline setup. We thank Maria Niculescu-Duvaz and Roger Wesson for their help with setting up DAMOCLES. This research is based in part on observations obtained at the international Gemini Observatory, a program of NSF’s NOIRLab, which is managed by the Association of Universities for Research in Astronomy (AURA) under a cooperative agreement with the National Science Foundation (NSF), on behalf of the Gemini Observatory partnership: the NSF (United States), National Research Council (Canada), Agencia Nacional de Investigación y Desarrollo (Chile), Ministerio de Ciencia, Tecnología e Innovación (Argentina), Ministério da Ciência, Tecnologia, Inovações e Comunicações (Brazil), and Korea Astronomy and Space Science Institute (Republic of Korea). A.P.R. and J.R. are in part supported by the NASA ADAP grant (80NSSC20K0449) for the study of SN dust. S.C.Y. and S.H.P. are supported by the National Research Foundation of Korea (NRF) grant (NRF-2019R1A2C2010885). Gemini telescope time is partly awarded by K-GMT of the Korea Astronomy and Space Science Institute (KASI). 

This work makes use of observations from the Las Cumbres Observatory global telescope network. The LCO team is supported by NSF grants AST-1911225 and AST-1911151. Research by S.V. and Y.D. is supported by NSF grant AST-2008108. R.K.T. is supported by the ÚNKP-22-4 New National Excellence Program of the Ministry for Culture and Innovation from the source of the National Research, Development and Innovation Fund, and the NKFIH/OTKA FK-134432 grant of the National Research, Development and Innovation (NRDI) Office of Hungary. S.B. is supported by grant RSF 19-12-00229 for the development of STELLA code and by RFBR  21-52-12032 in the studies of SNIc. A.V.F. is grateful for financial support from the Christopher R. Redlich Fund and numerous individual donors. The UCSC team is supported in part by the Gordon \& Betty Moore Foundation, the Heising-Simons Foundation, and by a fellowship from the David and Lucile Packard Foundation to R.J.F. A major upgrade of the Kast spectrograph on the Shane 3\,m telescope at Lick Observatory was made possible through generous gifts from the Heising-Simons Foundation as well as William and Marina Kast. Research at Lick Observatory is partially supported by a generous gift from Google. 
Some of the data presented herein were obtained at the W. M. Keck
Observatory, which is operated as a scientific partnership among the
California Institute of Technology, the University of California, and
NASA; the observatory was made possible by the generous financial
support of the W. M. Keck Foundation.

\newpage
\clearpage

\end{document}